# Hydrodynamic equations for a granular mixture from kinetic theory – fundamental considerations

**James W. Dufty**
*Department of Physics, University of Florida, Gainesville Fl 32611, USA*
**Aparna Baskaran**
*Department of Physics, Syracuse University, Syracuse NY 13244, USA*

**ABSTRACT**
In this review, a theoretical description is provided for the solid (granular) phase of the gas-solid flows that are the focus of this book. Emphasis is placed on the fundamental concepts involved in deriving a macroscopic hydrodynamic description for the granular material in terms of the hydrodynamic fields (species densities, flow velocity, and the granular temperature) from a prescribed "microscopic" interaction among the grains. To this end, the role of the interstitial gas phase, body forces such as gravity, and other coupling to the environment are suppressed and retained only via a possible non-conservative external force and implicit boundary conditions. The general notion of a kinetic equation is introduced to obtain macroscopic balance equations for the fields. Constitutive equations for the fluxes in these balance equations are obtained from special "normal" solutions to the kinetic equation, resulting in a closed set of hydrodynamic equations. This general constructive procedure is illustrated for the Boltzmann-Enskog kinetic equation describing a system of smooth, inelastic hard spheres. For weakly inhomogeneous fluid states the granular Navier-Stokes hydrodynamic equations are obtained, including exact integral equations for the transport coefficients. A method to obtain practical solutions to these integral equations is described. Finally, a brief discussion is given for hydrodynamics beyond the Navier-Stokes limitations.

## INTRODUCTION

Activated granular materials occur ubiquitously in nature and practical realizations in industry. Many of the phenomena occur on length scales that are large compared to the size of constituent particles (grains) and time scales long compared to the time between collisions among the grains. In this case a description of the system in terms of the values for hydrodynamic fields in cells containing many particles, analogous to molecular fluids, can apply for granular fluids. The hydrodynamic fields for molecular fluids are the densities associated with the globally conserved quantities. In the absence of reactions, these are the species densities, total momentum density, and the energy density. More commonly the momentum density is replaced by a corresponding flow field, and the energy density is replaced by a related temperature. The time dependence of such a macroscopic description (hydrodynamics) follows from the exact conservation equations for these fields, supplemented by "constitutive equations" providing a closed description in terms of the fields alone.

The key difference between granular and molecular fluids is that the former involves collisions between macroscopic grains. These collisions conserve momentum but dissipate energy since part of the kinetic energy of the grains goes into micro-deformations of the surface and exciting other internal modes of the grains. Even so, a hydrodynamic description for a fluid of grains can be given under appropriate conditions, following closely the approach developed in the context of molecular fluids, starting from the exact "balance equations" for the densities of interest. The objective of this chapter is to provide an overview of how general constitutive equations can be obtained from a fundamental basis in kinetic theory. The discussion does not make specific reference to a particular fluid state or kinetic theory. This overview is followed by a practical illustration for the special case of Navier-Stokes hydrodynamics for weakly non-uniform states, derived from the generalized Enskog kinetic theory (van Beijeren & Ernst 1973, 1979) extended to granular systems (Brey, Dufty & Santos 1997; see also Appendix A of Garzo, Dufty & Hrenya 2007). Extensive references to previous work on Navier-Stokes constitutive equations from Boltzmann and Enskog kinetic theories can be found in the review of Goldhirsch 2003, the text of Brilliantov & Poschel 2004, and in the recent articles Garzo, Dufty, & Hrenya 2007 and Garzo, Hrenya & Dufty 2007.

The balance equations are local identities expressing the change in hydrodynamic fields of a cell due to their fluxes through the boundaries of that cell and local sources within the cell. The central problem is to represent the fluxes and sources in terms of these hydrodynamic fields and their gradients. In many cases the form of these constitutive equations is known from experiments (e.g., Fick's diffusion law, Newton's viscosity law). An important advantage of kinetic theory as the basis for constitutive equations, in contrast to such phenomenology generalized from experiment, is that both quantitative and qualitative predictions follow as mathematical consequences of the theory. Thus the form of the hydrodynamic equations, the values of their parameters, and the validity conditions for applications are provided as one unit. In practice, most applications to granular fluids have focused on low density conditions and moderate densities at low dissipation (e.g., the Boltzmann and Enskog kinetic theories) (Jenkins & Mancini 1989, Jenkins 1998, Lun 1991). However, the approach emphasized here is more general and provides a means to describe quite general complex fluid states such as those that occur more generally for granular fluids. The aim of this chapter is to provide a pedagogical overview of the basis for hydrodynamics as arising from kinetic theories (for a similar analysis based on the low density Boltzmann equation see Dufty & Brey 2005). With this goal in mind, attention is restricted to the simplest case of smooth grains interacting through pair-wise additive short ranged interactions. Other important effects such as those due to the interstitial fluid phase, gravity etc. are included only at the level of an external body force acting on the grains, and are addressed briefly in the next section.

The layout of the chapter is as follows. Section 2 provides an overview of role of kinetic theory and hydrodynamics in the context of gas-solid flows, highlighting the advantages and limitations of each. In the next section the notion of a kinetic theory as a "mesoscopic" theory is introduced in its most general form. Next, the balance equations for the hydrodynamic fields are obtained from the kinetic theory with explicit expressions for the fluxes in terms of the solution to the kinetic equation. Finally, the notion of constitutive equations is introduced for special "normal" solutions to the kinetic equation. Together, the balance equations supplemented with the constitutive equations yield the closed hydrodynamic description of the fluid in terms of the local fields. These general considerations are formally exact and provide the basis for specialization to particular applications and practical approximations. The remainder of this chapter is then focused on the important case of states with small spatial and temporal variations, for which the Navier-Stokes hydrodynamic equations are obtained. The force law for particle-particle collisions in the kinetic theory is idealized to that of smooth, inelastic hard spheres, and the collision operator is specialized to a practical form (revised Enskog kinetic equation) appropriate for a wide range of space and time scales, and densities. As an important illustration, the normal solution is described for weakly inhomogeneous states as an expansion in the small spatial and temporal gradients, leading to the explicit constitutive equations and expressions for the associated transport coefficients. Finally, the need to go beyond Navier-Stokes hydrodynamics for many granular states is discussed. Applications of the Enskog kinetic theory to uniform shear flow is noted as an important example.

The scope of topics covered is quite broad and a complete citation of all the important literature is not practical. Instead, in many cases reference will be given to reviews in which extensive bibliographies appear. Generally, it is hoped that the material presented is self-contained in the sense that the logical presentation can be followed even though the full details of calculations are left implicit.

## CONTEXT: GAS-SOLID FLOWS

Before embarking on the details of a kinetic theory and its basis for hydrodynamics, it is useful to review the context of each in the description of the complex flows encountered in gas-solid systems (e.g. gas-fluidized beds). The complexity arises from a number of sources, e.g. gravitational field, geometry (boundary conditions), and formation of heterogeneous structures (macroscopic bubbles or high density clusters) (van Swaaij 1990, Gidaspow 1994). Since most features of interest occur at the laboratory scale, hydrodynamics has been a primary tool in attempts to model gas-solid flows (Jackson 2000, Kuipers, Hoomans & van Swaaij 1998). On this large scale, both gas and particle subsystems are described by Navier-Stokes continuum equations with sources coupling the two self-consistently (the Two Fluid Model, Anderson & Jackson 1967). However, the parameters of these equations such as the particle-gas force and transport properties must be supplied phenomenologically and their detailed forms can make a significant difference in the flows predicted. To overcome this limitation, a more detailed description on smaller length scales is required.

One approach is to describe the particle dynamics by numerical simulation of the associated Newton's equations of motion, while retaining a hydrodynamic description for the gas. This is the Discrete Particle Model (also referred to as the Discrete Element Method) (Hoomans, Kuipers, Briels, van Swaaij 1996; Deen, van Sint Annaland, van der Hoef & Kuipers 2007, Zhu, Zhou, Yang & Yu 2007 ). This provides a detailed and accurate description of the particle trajectories for given forces. In addition to gravity, pressure gradients, particle-particle and particle-wall forces, an essential force in Newton's equation is that of the gas on the fluid particle. Implementation of the DPM can lead to different results depending on the nature and treatment of the forces on the particle (Feng & Yu 2004, Leboriero, Joseph & Hrenya 2008, van Wachem et al., 2007). Although this drag force on the particle is localized at the surface of the particle, it can depend on the details of the gas fluid state including the indirect influence of other particles on this state. For example, it can be different for dense monodisperse and polydisperse gas-solid systems (van der Hoef, Beetstra & Kuipers 2005). Recently, accurate modeling of the gas-particle drag force has been possible using lattice gas Boltzmann methods that allow an accurate simulation of the gas on a lattice smaller than the particle size, accounting for both the momentum transfer to the particles and incorporating details of the boundary conditions. In this way lattice gas Boltzmann simulations on the smallest scale provide the needed input for DPM on the mesoscopic scale (Hill, Koch & Ladd 2001; Benyahia, Syamlal & O'Brien 2006; Yin & Sundaresan 2009; van der Hoef, van Sint Annaland & Kuipers 2004).

A kinetic theory description for the particles provides an alternative to the DPM on the same scale of the particle positions and velocities. There are two main advantages of kinetic theory. First, it does not have the practical limitations of discrete particle simulations to small (compared to laboratory) systems of particles. Second, as described below, the transition to hydrodynamics and identification of its parameters is straightforward from kinetic theory, but very much less so for DPM. It requires the same input forces, both for collisions and for coupling to the gas phase (still described by Navier-Stokes equations), and so can benefit from the recent developments for DPM. On the other hand, for very dense clusters and glassy structures the form of the particle-particle collisions in kinetic theory is only known semi-phenomenologically at this point. Applications of kinetic theory to gas-solid flows are mainly in the context of providing the form and parameters of the two fluid model (see however Minier & Peirano 2001). Early examples of this approach include Sinclair & Jackson 1989, and Koch (1990); for a review and references see Gidaspow, Jung, & Singh 2004. Recent improvements in the kinetic theory and its systematic application for the normal solution have led to a more accurate solid phase hydrodynamics, as described below. The coupled sets of continuum equations in the two fluid model then constitute a problem in computational fluid dynamics, often including additional assumptions for the gas phase to describe turbulent conditions, as described elsewhere in this book.

The derivation of hydrodynamics (specifically, constitutive equations) generally entails necessary conditions, e.g., sufficiently small Knudsen numbers for Navier-Stokes hydrodynamics. As emphasized below, the term "hydrodynamics" includes more general fluid states with correspondingly more complex constitutive equations. In any case, such closures have associated validity conditions that must be checked before application to a given problem. The complication arising from the particle-gas drag force can affect these validity conditions significantly. The kinetic equation remains valid more generally, just as DPM, but the possibility of a hydrodynamic description might be precluded. Under these conditions, the particle hydrodynamic description must be replaced with a more general solution to the kinetic equation itself.

The objective of the current chapter is focused on the *method* of deriving systematically hydrodynamic equations from a given kinetic equation. The specific kinetic equation considered and the complexity of the resulting hydrodynamic description can depend on the flow conditions, geometry, degree of heterogeneity, etc. and such applications are the subject of other chapters in this book. The illustration of this method given here is for an ideal granular fluid of inelastic hard spheres described by the generalized Enskog equation. It is expected that this is a good compromise between accuracy and practical utility, not limited by conditions of Knudsen number, Reynolds number, density heterogeneity, or geometry. In most respects it has both the generality of DPM and the advantages of kinetic theory. The application of that kinetic theory to Navier-Stokes hydrodynamics described here, however, has the additional more severe limitations to small Knudsen numbers. While this is the most common hydrodynamics currently in use for the two fluid description of gas-solid flows, it is clear from the derivation here that failure should be expected for many conditions of interest (e.g., bubbles, plugs, rheology). Nevertheless, the systematic derivation of Navier-Stokes hydrodynamics provides accurate results under its validity conditions for both the form of those equations as well as quantitative values for the transport coefficients. For example, the physical mechanisms that govern polydisperse mixing and separation processes (Duran Rajchenbach and Clement 1993) are not well understood yet; the recent results described here have been applied to a controlled, quantitative means to study one of those mechanisms, thermal diffusion (Garzo 2009).

## KINETIC THEORY AS A BASIS FOR HYDRODYNAMICS
### Kinetic Theory

Consider a mixture of *s* species of smooth spherical particles with masses $\{m_i; i = 1..s\}$. Their sizes and material composition can be suppressed at this point as they enter only through the force laws for the particle-particle and particle-gas interactions. These are taken to be short ranged (compared to relevant cell sizes for the macroscopic description), conserve momentum, but dissipate energy. The hydrodynamic fields of interest describe a few densities at each spatial point in the system. A more complete mesoscopic description is given by the distribution of particles in the six dimensional phase space defined by the points $\mathbf{r}, \mathbf{v}$, where $\mathbf{r}$ is the position and $\mathbf{v}$ is the velocity of a particle. For a system with *s* different species, there is a set of distribution functions for all the species $\{f_i(\mathbf{r}, \mathbf{v}; t); i = 1, ..., s\}$. If these functions are normalized to unity, then each $f_i(\mathbf{r}, \mathbf{v}; t)$ is the probability density of finding a particle of species *i* at position $\mathbf{r}$ with velocity $\mathbf{v}$ at the time $t$. In the following, the normalization is chosen instead to be the species densities so the interpretation is that of a number density of species *i* at $\mathbf{r}, \mathbf{v}, t$.

The species densities $\{n_i(\mathbf{r}, t)\}$, energy density $e(\mathbf{r}, t)$, and momentum density $\mathbf{p}(\mathbf{r}, t)$ are defined in terms of the distribution functions by

$$n_i(\mathbf{r},t) = \int d\mathbf{v} f_i(\mathbf{r},\mathbf{v};t), \qquad i = 1,\ldots,s \tag{1}$$

$$e(\mathbf{r},t) = \int d\mathbf{v}\frac{1}{2}m_i v^2 f_i(\mathbf{r},\mathbf{v};t) + \frac{1}{2}\sum_{i,j=1}^{s}\int d\mathbf{v}d\mathbf{v}'\,d\mathbf{r}'V_{ij}(\mathbf{r}-\mathbf{r}')f_{ij}(\mathbf{r},\mathbf{v};\mathbf{r}',\mathbf{v}';t), \tag{2}$$

$$\mathbf{p}(\mathbf{r},t) = \int d\mathbf{v}\,m_i\mathbf{v}f_i(\mathbf{r},\mathbf{v};t). \tag{3}$$

Here, $V_{ij}(\mathbf{r}-\mathbf{r}')$ is the potential energy for a pair of particles associated with the conservative part of the force between them, and $f_{ij}$ is the joint distribution function for two particles. It is usual to introduce a flow field $\mathbf{U}$ in place of the momentum density according to

$$\mathbf{p}(\mathbf{r},t) \equiv \rho(\mathbf{r},t)\,\mathbf{U}(\mathbf{r},t). \tag{4}$$

Similarly, the energy in the local rest frame is represented by a corresponding kinetic temperature $T$

$$e(\mathbf{r},t) \equiv \frac{3}{2}n(\mathbf{r},t)T(\mathbf{r},t) + \frac{1}{2}\rho(\mathbf{r},t)U^2(\mathbf{r},t), \qquad n(\mathbf{r},t) = \sum_i n_i(\mathbf{r},t). \tag{5}$$

In the following, $\{n_i(\mathbf{r},t)\}$, $T(\mathbf{r},t)$, and $\mathbf{U}(\mathbf{r},t)$ will be referred to as the independent hydrodynamic fields.

These definitions express the dynamics of the hydrodynamic fields in terms of the more fundamental dynamics of the distribution functions. Their evolution follows exactly from Newton's equations (Ernst 2000)

$$\begin{aligned}(\partial_t + \mathbf{v}_1\cdot\nabla_{\mathbf{r}_1})f_i(\mathbf{r}_1,\mathbf{v}_1;t) &+ \nabla_{\mathbf{v}_1}\cdot(m_i^{-1}\mathbf{F}_{0i}(\mathbf{r}_1,\mathbf{v}_1)f_i(\mathbf{r}_1,v_1;t))\\ &= -\nabla_{\mathbf{v}_1}\cdot\sum_{j=1}^{s}\int d\mathbf{r}_2 d\mathbf{v}_2 m_i^{-1}\mathbf{F}_{ij}(\mathbf{r}_{12},\mathbf{v}_{12})f_{ij}(\mathbf{r}_1,\mathbf{v}_1;\mathbf{r}_2,\mathbf{v}_2;t),\end{aligned} \tag{6}$$

The left side describes the evolution of a distribution without interparticle interactions, in the presence of a (possibly non-conservative) external force $\mathbf{F}_{0i}$ due to confinement, gravity, and coupling to the gas (in gas-solid flows). The right side represents the changes in the distribution for species $i$ due to a force from a particle of species $j$. The occurrence of that other particle at any point with any velocity is given by the joint distribution for two particles $f_{ij}(\mathbf{r}_1,\mathbf{v}_1;\mathbf{r}_2,\mathbf{v}_2;t)$. The force law $\mathbf{F}_{ij}$ is chosen to depend on the relative distances $\mathbf{r}_{12} = \mathbf{r}_1 - \mathbf{r}_2$ and relative velocities $\mathbf{v}_{12} = \mathbf{v}_1 - \mathbf{v}_2$, representing a central force conserving total momentum, but not conserving energy. Model interactions that belong to this class include the Hertzian contact force model (Campbell 1990)

$$\mathbf{F}_{ij}(\mathbf{r},\mathbf{v}) = \hat{\mathbf{r}}\,\Theta(\sigma-r)\left(a_{ij}(\sigma-r)^{3/2} - c_{ij}\,\hat{\mathbf{r}}\cdot\mathbf{v}\right), \tag{7}$$

where the first term describes response to elastic deformation and the second term describes the dissipation of energy during this deformation. In general there is an additional tangential component of the force as well, describing the roughness of the grains. More general non–central force laws representing the shape of the particles can be included, but will not be considered here. Below, a limiting form of eqn. (7) representing smooth, inelastic hard spheres with constant or velocity dependent coefficient of restitution will also be introduced. The rest of the chapter focused on developing the theory in the context of these two model interactions.

Equation (6) is exact and applicable to quite general state conditions. However, it couples the distribution $f_i(\mathbf{r}_1, \mathbf{v}_1; t)$ to the two particle distributions $f_{ij}(\mathbf{r}_1, \mathbf{v}_1; \mathbf{r}_2, \mathbf{v}_2; t)$. These obey similar equations but are coupled to still higher order multi-particle distributions. The resulting set of equations is known as the Born, Bogoliubov, Green, Kirkwood, Yvon (BBGKY) hierarchy (McLennan 1989; Résibois & De Leener 1977; Ferziger & Kaper 1972). In contrast, a kinetic theory is a closed equation for the set of single particle functions $\{f_i(\mathbf{r}_1, \mathbf{v}_1; t)\}$ alone. Such closed equations result from eqn. (6) if the two particle functions can be expressed as functionals of the one particle functions

$$f_{ij}(\mathbf{r}_1, \mathbf{v}_1; \mathbf{r}_2, \mathbf{v}_2; t) = \mathscr{F}_{ij}(\mathbf{r}_1, \mathbf{v}_1; \mathbf{r}_2, \mathbf{v}_2 / \{f_k(t)\}). \tag{8}$$

If such a functional can be found, then (6) becomes a closed kinetic equation

$$(\partial_t + \mathbf{v}_1 \cdot \nabla_{\mathbf{r}_1}) f_i(\mathbf{r}_1, \mathbf{v}_1; t) + \nabla_{\mathbf{v}_1} \cdot (m_i^{-1} \mathbf{F}_{0i}(\mathbf{r}_1, \mathbf{v}_1) f_i(\mathbf{r}_1, \mathbf{v}_1; t)) = C_i(\mathbf{r}_1, \mathbf{v}_1 \mid \{f_k(t)\}), \tag{9}$$

with the "collision operator"

$$C_i(\mathbf{r}_1, \mathbf{v}_1 \mid \{f_k(t)\}) = -\nabla_{\mathbf{v}_1} \cdot \sum_{j=1}^{s} \int d\mathbf{r}_2 d\mathbf{v}_2 m_i^{-1} \mathbf{F}_{ij}(\mathbf{r}_{12}, \mathbf{v}_{12}) \mathscr{F}_{ij}(\mathbf{r}_1, \mathbf{v}_1; \mathbf{r}_2, \mathbf{v}_2 \mid \{f_k(t)\}). \tag{10}$$

*This constitutes the most general definition of a kinetic theory.*

The discovery of the functional (i.e., a non-local dependence on the fields at all points) in eqn. (8) and the corresponding collision operator is the point at which the difficult many-body problem is confronted. It is common to all theoretical descriptions of macroscopic systems and there is a long history for molecular fluids (Bogoliubov 1962; Cohen 1962). Some analyses are systematic when there is a small parameter. For example, at low density the granular Boltzmann equation can be recovered in this way (Dufty 2001). More generally, there is a mixture of analysis and phenomenology combined with feedback from comparison of predictions with experiments. This is an area of active current investigation for dense granular systems. It will not be discussed further here beyond emphasizing that this notion of a kinetic theory does not preclude the description of quite complex granular states, far outside the limitations of Boltzmann kinetic theory.

## Macroscopic Balance Equations

The macroscopic balance equations are those for the time derivatives of the hydrodynamic fields. They follow directly from their definitions above in terms of integrals over $f_i(\mathbf{r}, \mathbf{v}; t)$ and the kinetic equation. The balance equations for the number densities, energy density, and momentum are obtained in their familiar forms

$$\partial_t n_i(\mathbf{r}, t) + m_i^{-1} \nabla \cdot \mathbf{j}_i(\mathbf{r}, t) = 0, \tag{11}$$

$$\partial_t e(\mathbf{r}, t) + \nabla \cdot \mathbf{s}(\mathbf{r}, t) = -w(\mathbf{r}, t) + \sum_{i=1}^{s} \int d\mathbf{v} \mathbf{F}_{0i}(\mathbf{r}, \mathbf{v}) \cdot \mathbf{v} f_i(\mathbf{r}, \mathbf{v}, t), \tag{12}$$

$$\partial_t p_\beta(\mathbf{r},t) + \nabla_\gamma t_{\gamma\beta}(\mathbf{r},t) = \sum_{i=1}^s \int d\mathbf{v} F_{0i\beta}(\mathbf{r},\mathbf{v}) f_i(\mathbf{r},\mathbf{v},t). \tag{13}$$

The left sides are the expected forms of a time derivative for the density plus the divergence of a flux. The right sides describe the sources and external forces. For example $w(\mathbf{r},t)$ is the energy density loss rate due to the non-conservative collisions among particles. In obtaining these expressions from the kinetic equation, the mass fluxes $\{\mathbf{j}_i\}$, energy flux $\mathbf{s}$, and momentum flux $t_{\gamma\beta}$ are obtained as explicit linear integrals over the $\{f_k(t)\}$ and $\{\mathcal{F}_{ij}\}$. To describe them, the contributions from pure convection are first identified,

$$\mathbf{j}_i = \mathbf{j}_{0i}(\mathbf{r},t) + \rho_i(\mathbf{r},t)\mathbf{U}(\mathbf{r},t), \tag{14}$$

$$s_\beta(\mathbf{r},t) = q_\beta(\mathbf{r},t) + \left(\frac{3}{2}n(\mathbf{r},t)T(\mathbf{r},t) + \frac{1}{2}\rho(\mathbf{r},t)U^2(\mathbf{r},t)\right)U_\beta(\mathbf{r},t) + P_{\beta\gamma}(\mathbf{r},t)U_\gamma(\mathbf{r},t), \tag{15}$$

$$t_{\gamma\beta}(\mathbf{r},t) = P_{\beta\gamma}(\mathbf{r},t) + \rho(\mathbf{r},t)U_\beta(\mathbf{r},t)U_\gamma(\mathbf{r},t). \tag{16}$$

The first terms on the right sides represent the corresponding flux in the local rest frame for each cell: the diffusion fluxes $\mathbf{j}_{0i}(\mathbf{r},t)$, the heat flux $\mathbf{q}(\mathbf{r},t)$, and the pressure tensor $P_{\beta\gamma}(\mathbf{r},t)$. Their explicit forms from the kinetic theory are (using the Hertzian force model (7) as an illustration of the force law (for the case of hard spheres see Lutsko 2004))

$$\mathbf{j}_{0i}(\mathbf{r}_1,t) \equiv m_i \int d\mathbf{v}_1 \mathbf{V}_1 f_i(\mathbf{r}_1,\mathbf{v}_1,t), \tag{17}$$

$$\mathbf{q}(\mathbf{r}_1,t) \equiv \sum_{i=1}^s \int d\mathbf{v}_1 \frac{1}{2} m_i V_1^2 f_i(\mathbf{r}_1,\mathbf{v}_1,t)$$
$$+ \frac{1}{2}\sum_{i,j=1}^s \int d\mathbf{v}_1 \int d\mathbf{v}_2 \int_0^1 d\lambda \lambda^3 \int d\hat{\sigma} \left[(\mathbf{G}_{ij}\cdot\hat{\sigma})a_{ij}(\sigma-\lambda)^{3/2} + (\mu_{ij}-\mu_{ji})c_{ij}(\mathbf{g}\cdot\hat{\sigma})^2\right]$$
$$\times \int_0^1 d\kappa \mathcal{F}_{ij}(\mathbf{r}_1 - (1-\kappa)\lambda\hat{\sigma},\mathbf{v}_1;\mathbf{r}_1 + \kappa\lambda\hat{\sigma},\mathbf{v}_2 / \{f_k(t)\}), \tag{18}$$

$$P_{\gamma\beta}(\mathbf{r}_1,t) = \sum_{i=1}^s \int d\mathbf{v}_1 \frac{1}{2} m_i V_{1\beta} V_{1\gamma} f_i(\mathbf{r}_1,\mathbf{v}_1,t)$$
$$+ \frac{1}{2}\sum_{i,j=1}^s \int d\mathbf{v}_1 \int d\mathbf{v}_2 \int_0^1 d\lambda \lambda^3 \int d\hat{\sigma}\hat{\sigma}_\gamma \hat{\sigma}_\beta \left[a_{ij}(\sigma-\lambda)^{3/2} + c_{ij}(\mathbf{g}\cdot\hat{\sigma})\right]$$
$$\times \int_0^1 d\kappa \mathcal{F}_{ij}(\mathbf{r}_1 - (1-\kappa)\lambda\hat{\sigma},\mathbf{v}_1;\mathbf{r}_1 + \kappa\lambda\hat{\sigma},\mathbf{v}_2 / \{f_k(t)\}), \tag{19}$$

where $\mathbf{V}_1 = \mathbf{v}_1 - \mathbf{U}(\mathbf{r},t)$ is the velocity in the local rest frame, $\mathbf{G}_{ij} = \mu_{ij}\mathbf{V}_1 + \mu_{ji}\mathbf{V}_2$ is the center of mass velocity of the two colliding particles where $\mu_{ij} = m_i/(m_i + m_j)$ is the reduced mass of species $i$ with respect to species $j$ and $\mathbf{g} = \mathbf{v}_1 - \mathbf{v}_2$ is the relative velocity of the two particles. Similarly, the energy loss rate is found to be

$$w(\mathbf{r},t) = \sum_{i,j=1}^{s}\int d\mathbf{v}_1 \int d\mathbf{v}_2 \int_0^1 d\lambda \lambda^2 \int d\hat{\sigma} c_{ij}(\mathbf{g}\cdot\hat{\sigma})^2 \mathcal{F}_{ij}(\mathbf{r}_1,\mathbf{v}_1;\mathbf{r}_1 - \lambda\hat{\sigma},\mathbf{v}_2/\{f_k(t)\}). \tag{20}$$

The first terms on the right sides are the fluxes due simply to the motion of the particles (kinetic fluxes), while the second terms in eqns. (18), (19), and (20) are due to the forces between particles (collisional transfer).
Substituting eqns. (14) - (16) into (11) - (13) gives the balance equations in the desired form

$$D_t n_i + n_i \nabla \cdot \mathbf{U} + m_i^{-1} \nabla \cdot \mathbf{j}_{0i} = 0, \tag{21}$$

$$\frac{3}{2}nD_t T + P_{\gamma\beta}\partial_{r_\gamma} U_\beta + \nabla \cdot \mathbf{q} - \frac{3}{2}T\sum_{i=1}^{s} m_i^{-1}\nabla \cdot \mathbf{j}_{0i} = -w + \sum_{i=1}^{s}\int d\mathbf{v}\mathbf{F}_{0i}\cdot(\mathbf{v}-\mathbf{U})f_i, \tag{22}$$

$$\rho D_t U_\beta + \partial_{r_\gamma} P_{\gamma\beta} = \sum_{i=1}^{s} n_i(\mathbf{r},t)\int d\mathbf{v} F_{0i\beta} f_i, \tag{23}$$

where $D_t = \partial_t + \mathbf{U}\cdot\nabla$ is the material derivative. *These balance equations for the hydrodynamic fields are an exact consequence of Newton's equations.* They have the same form as those for a molecular fluid (McLennan 1989), except for the source $w$ in the temperature equation due to non-conservative forces. The fluxes $\{\mathbf{j}_{0i}\}$, $\mathbf{q}$, $P_{\gamma\beta}$, and the energy source $w$ are not given in terms of the fields so these equations are not "closed", i.e. they are not self-determined by the fields themselves. The terms involving the averages of the velocity dependent external force are also required. However, all of these unknowns are given in terms of the solution to the kinetic equation through eqns. (17) - (20) which provides the controlled means for discovering the appropriate forms of these fluxes in terms of the fields. In this general context, the form of the balance equations is independent of the specific particle-particle interaction and the coupling to its environment.

## "Normal" States and Hydrodynamics

A true macroscopic description is obtained when eqns. (21) - (23) can be solved for the fields from their given initial and boundary conditions. This requires a "closure" whereby the fluxes and source $\{\mathbf{j}_{0i}\}$, $\mathbf{q}$, $P_{\gamma\beta}$, and $w$ are expressed as functionals of the hydrodynamic fields through constitutive equations. This is obtained directly from eqns. (17) - (20) by constructing solutions to the kinetic equations that are expressed as functionals of the fields. Such solutions are called normal solutions (McLennan 1989). They are characterized by the fact that all space and time dependence of the $\{f_i(\mathbf{r},\mathbf{v};t)\}$ occurs only through functionals of the hydrodynamic fields (denoted in the following collectively by $\{y_\alpha(t)\}$)

$$f_i(\mathbf{r},\mathbf{v};t) \rightarrow f_i(\mathbf{r},\mathbf{v};\{y_\alpha(t)\}), \tag{24}$$

so that, for example, the space and time derivatives become

$$\begin{pmatrix} \partial_t f_i(\mathbf{v}/\{y_\alpha(\mathbf{r},t)\}) \\ \nabla f_i(\mathbf{v}/\{y_\alpha(\mathbf{r},t)\}) \end{pmatrix} = \int d\mathbf{r}' \sum_\beta \frac{\delta f_i(\mathbf{r},\mathbf{v}/\{y_\alpha(t)\})}{\delta y_\beta(\mathbf{r}',t)} \begin{pmatrix} \partial_t y_\beta(\mathbf{r}',t) \\ \nabla' y_\beta(\mathbf{r}',t) \end{pmatrix}. \tag{25}$$

Generally, an initial preparation of the system will not have this normal form. However, as for molecular fluids, it is expected that there is a short "kinetic" stage during which particles in each cell have their velocities relax toward a universal form (e.g., Maxwellian for molecular fluid), but with values for the hydrodynamic fields different for each cell. On a longer time scale the normal form can be supported.

For such solutions the functional $\mathscr{F}_{ij}$ becomes normal as well,

$$\mathscr{F}_{ij}(\mathbf{r}_1,\mathbf{v}_1;\mathbf{r}_2,\mathbf{v}_2 \mid f_k(\{y_\alpha(t)\})) \to \mathscr{G}_{ij}(\mathbf{r}_1,\mathbf{v}_1;\mathbf{r}_2,\mathbf{v}_2 \mid \{y_\alpha(t)\}). \tag{26}$$

Consider some arbitrary property $A_i(\mathbf{r},t)$ for species $i$ defined as the average of $a(\mathbf{v})$. For normal states, all space and time dependence of this property occurs as a functional of the hydrodynamic fields

$$A_i(\mathbf{r},t) = \int d\mathbf{v} a(\mathbf{v}) f_i(\mathbf{r},\mathbf{v} \mid \{y_\alpha(t)\}) = A_i(\mathbf{r} \mid \{y_\alpha(t)\}).$$

In this way the averages defining the fluxes and energy source, eqns. (17) - (20), become the desired constitutive equations, which for the Hertzian model eqn. (7) are

$$\mathbf{j}_{0i}(\mathbf{r} \mid \{y_\alpha(t)\}) = m_i \int d\mathbf{v} \mathbf{V} f_i(\mathbf{r},\mathbf{v} \mid \{y_\alpha(t)\}), \tag{27}$$

$$\mathbf{q}(\mathbf{r} \mid \{y_\alpha(t)\}) = \sum_{i=1}^s \int d\mathbf{v} \frac{1}{2} m_i V^2 \mathbf{V} f_i(\mathbf{r},\mathbf{v} \mid \{y_\alpha(t)\})$$

$$+ \sum_{i,j=1}^s \int d\mathbf{v}_1 \int d\mathbf{v}_2 \int_0^\sigma d\lambda \lambda^3 \int d\hat{\sigma} \left[ (\mathbf{G}_{ij} \cdot \hat{\sigma}) a_{ij}(\sigma-\lambda)^{3/2} + (\mu_{ij} - \mu_{ji}) c_{ij} (\mathbf{g} \cdot \hat{\sigma})^2 \right]$$

$$\times \int_0^1 d\kappa \mathscr{G}_{ij}(\mathbf{r}-(1-\kappa)\lambda\hat{\sigma},\mathbf{v}_1;\mathbf{r}+\kappa\lambda\hat{\sigma},\mathbf{v}_2 \mid \{y_\alpha(t)\}), \tag{28}$$

$$P_{\gamma\beta}(\mathbf{r} \mid \{y_\alpha(t)\}) = \sum_{i=1}^s \int d\mathbf{v} m_i V_\beta V_\gamma f_i(\mathbf{r},\mathbf{v} \mid \{y_\alpha(t)\})$$

$$+ \sum_{i,j=1}^s \int d\mathbf{v}_1 \int d\mathbf{v}_2 \int_0^\sigma d\lambda \lambda^3 \int d\hat{\sigma} \hat{\sigma}_\gamma \hat{\sigma}_\beta \left( a_{ij}(\sigma-\lambda)^{3/2} + c_{ij}(\mathbf{g} \cdot \hat{\sigma}) \right)$$

$$\times \int_0^1 d\kappa \mathscr{G}_{ij}(\mathbf{r}-(1-\kappa)\lambda\hat{\sigma},\mathbf{v}_1;\mathbf{r}+\kappa\lambda\hat{\sigma},\mathbf{v}_2 \mid \{y_\alpha(t)\}), \tag{29}$$

and

$$w(\mathbf{r} \mid \{y_\alpha(t)\}) = \sum_{i,j=1}^s \int d\mathbf{v}_1 \int d\mathbf{v}_2 \int_0^\sigma d\lambda \lambda^3 \int d\hat{\sigma} c_{ij} (\mathbf{g} \cdot \hat{\sigma})^2 \mathscr{G}_{ij}(\mathbf{r},\mathbf{v}_1;\mathbf{r}-\lambda\hat{\sigma},\mathbf{v}_2 \mid \{y_\alpha(t)\}). \tag{30}$$

*The balance equations (21) - (23) together with the constitutive equations (27) - (30) constitute the most general definition of hydrodynamics for a molecular or granular fluid.*

It is appropriate at this point to pause and discuss the choice of independent hydrodynamic fields. The fundamental idea is that such fields should represent the dominant dynamics on large space and time scales. If they are local conserved densities this property is assured since they ultimately approach constants as the system becomes uniform. Hence, for a molecular fluid the species densities, momentum density (or flow velocity), and the energy density (or temperature) are the clear choices. Note that the species energy densities or partial temperatures are not conserved and are therefore not appropriate choices for independent hydrodynamic fields. For granular fluids, the species densities and momentum density are still conserved and are proper choices for fields. However, now the total energy density (temperature), are not conserved due to the inelastic collisions, and it is not clear that its dynamics should dominate other kinetic modes on the long time scale. For the moment it will be assumed that this dominance still applies, and further discussion is provided below. In any case, just as for molecular fluids, the partial temperatures are not independent fields. For notational simplicity it may be useful to introduce partial temperatures through the definition,

$$\frac{3}{2} n_i T_i \equiv \int d\mathbf{v} \frac{1}{2} m_i V^2 f_i(\mathbf{r}, \mathbf{v} | \{y_\alpha(t)\}). \tag{31}$$

However, these should be viewed simply as measures of the second moments of the species distributions and, as the last equality emphasizes, these partial temperatures are determined as functions of the hydrodynamic fields chosen here. The definition of $T$ in eqn. (5) then implies the identity

$$T = \frac{\sum_i n_i T_i(\{n_i\}, T)}{\sum_i n_i}. \tag{32}$$

In general, for mechanically different species, the partial temperatures are all different and not equal to $T$ (Garzo & Dufty 1999).

This concludes the characterization of the formal basis of macroscopic hydrodynamics of a multicomponent granular fluid as arising from a more mesoscopic description of the system given by a general kinetic theory of the associated one particle distribution functions.

## Uniform Fluid Hydrodynamics

Solutions to the hydrodynamic equations cannot be addressed until the details of the constitutive equations are specified and suitable initial and boundary values given. An exception is the simplest case of an isolated, uniform fluid. The spatial variations of the fields vanish, and with stationary uniform boundary conditions and no external forces a molecular fluid would be in its equilibrium state. For a granular fluid, however, eqns. (21) - (23) become

$$\partial_t n_i = 0 = \partial_t U_\beta, \quad \partial_t T = -\zeta(\{n_i\}, T)T, \tag{33}$$

where the "cooling rate" $\zeta$ has been introduced in place of the energy loss rate

$$\zeta(\{n_i\}, T) \equiv \frac{2w(\{n_i\}, T)}{3nT}. \tag{34}$$

The solution represents a uniform fluid at rest with a monotonically decreasing temperature, and is known as the homogeneous cooling state (HCS). This state was first discussed by Haff (Haff 1983), and subsequently studied via low density kinetic theory (Brey, Ruiz-Montero & Cubero 1996; van Noije & Ernst 1998), and more generally by molecular dynamics simulations (Deltour & Barrat 1997; Goldhirsch, Tan & Zanetti 1993; McNamara & Young 1996). The corresponding normal solution to the kinetic equation in this case is the homogeneous cooling solution, obtained by substituting the normal form eqn. (24) for a homogeneous state into the kinetic equation (9)

$$-\zeta(\{n_i\},T)T\partial_T f_i(\mathbf{v}/\{y_\alpha(t)\}) = C_i(\mathbf{v}/\{f_i(\{y_\alpha(t)\})\}). \tag{35}$$

As will become clear in the next section, a *local* HCS for each cell with the local values for the hydrodynamic fields is the reference state about which the hydrodynamic description of an isolated granular fluid is constructed. In this sense, the local HCS plays the same role for a granular fluid that the local equilibrium state plays for the molecular fluids, and hence the HCS is an important state to study and characterize. Further discussion of the HCS is given below in the specific context of the hard sphere granular fluid.

## NAVIER STOKES HYDRODYNAMICS FOR THE HARD SPHERE FLUID

The analysis of the above section shows that the derivation of hydrodynamics from kinetic theory has a very general context. There are two main difficulties in implementing this generic prescription. The first is the determination of the functional $\mathscr{F}_{ij}(\mathbf{r}_1,\mathbf{v}_1;\mathbf{r}_2,\mathbf{v}_2/\{f_k(\{y_\alpha(t)\})\})$ in eqn. (8) which provides the kinetic equation. The second is finding the normal solution to the given kinetic equation to obtain the constitutive equations. In the remainder of this chapter, the application of a normal solution is illustrated for the special case of a granular fluid modeled as a mixture of smooth, inelastic hard spheres. A practical kinetic theory in this case, applicable over a wide range of densities, is given by the generalized Enskog equation (van Beijeren & Ernst 1973, 1979; Brey, Dufty & Santos 1997; Garzó, Dufty, & Hrenya 2007). The normal solution for this kinetic equation is obtained by the Chapman-Enskog method (Ferziger & Kaper 1972; Brey, Dufty, Kim, Santos 1998; Garzó and Dufty 1999) for states with small spatial variations of the hydrodynamic fields over distances of the order of the mean free path. The resulting hydrodynamic equations are partial differential equations with spatial derivatives up to degree two, known as the Navier-Stokes equations for a granular fluid. Recent applications of hydrodynamics and kinetic theory for granular fluids in the context of the Enskog equation and its low density limit, the Boltzmann equation, can be found in Pöschel & Luding 2001, Pöschel & Brilliantov, 2003, and Brilliantov & Pöschel, 2004. The presentation here for the generalized Enskog equation follows the recent work of Garzo, Dufty, & Hrenya 2007.

### Enskog Kinetic Theory

As noted above, the derivation of a kinetic equation requires confrontation of the difficult many-body problem in nonequilibrium statistical mechanics. The analysis for molecular fluids is most complete for the idealized force law of hard, elastic spheres. This is a realistic quantitative model as well because the real short ranged repulsion is $\approx (\sigma/r)^p$, where $\sigma$ is the particle–particle force range (particle size), and $p$ is an integer. Since $p$ is large for repulsive interactions the hard sphere limit $(p \to \infty)$ is a good idealization. A detailed analysis of the hard sphere limit for molecular fluids is given in (Dufty 2002; Dufty & Ernst 2004). Similarly for granular fluids, the repulsive part of the force in (7) is characterized by parameters $a_{ij}$ that determine the rigidity of the interaction. For rigid particles the fractional compression is small and the velocity changes occur in a very short time. During this time, however, some energy is lost so the normal component of the asymptotic relative velocity is decreased. This type of collision is captured by the idealization of inelastic hard spheres that replace the short collision time by an instantaneous collision and the energy loss is captured by a single parameter for each pair of species, namely a coefficient of restitution $\alpha_{ij}$. For this model interaction, the velocities of a pair of particles undergoing a collision event change instantaneously according to

$$\mathbf{v}'_1 = \mathbf{v}_1 - \frac{m_j}{m_i + m_j}(1+\alpha_{ij})(\hat{\sigma}\cdot\mathbf{g}_{12})\hat{\sigma}, \quad \mathbf{v}'_2 = \mathbf{v}_2 + \frac{m_i}{m_i + m_j}(1+\alpha_{ij})(\hat{\sigma}\cdot\mathbf{g}_{12})\hat{\sigma}. \tag{36}$$

The prime denotes the velocities after collisions, $\mathbf{g}_{12} \equiv \mathbf{v}_1 - \mathbf{v}_2$, and $\hat{\sigma}$ is a unit vector directed along the line of the centers from 2 to 1. The particles are characterized by their masses and diameters $\{m_i, \sigma_i\}$, and a set of restitution coefficients for collisions among the same and different species $\{\alpha_{ij}\}$. The change in kinetic energy for the pair in eqn. (6) is $\Delta E_{ij} = -(1-\alpha_{ij}^2)\frac{m_i \mu_{ji}}{4}(\mathbf{g}_{12}\cdot\hat{\sigma})^2$. Thus the collisions are elastic for $\alpha_{ij} = 1$ and inelastic for $0 < \alpha_{ij} < 1$. It is easily verified that the total momentum is conserved for these collisions in both cases. In general these coefficients of restitution should depend on the normal component of the relative velocity for the pair. However, for sufficiently activated particles it is possible to consider a simpler model for which the $\alpha_{ij}$ are constants. For notational simplicity below, it is useful to introduce a substitution operator $b_{ij}$ that changes the velocities of a function according to the rule eqn. (36)

$$X(\mathbf{v}'_1, \mathbf{v}'_2) = b_{ij} X(\mathbf{v}_1, \mathbf{v}_2). \tag{37}$$

The form eqn. (36) implies no change in the tangential components of the velocities (perpendicular to $\hat{\sigma}$) and so represents smooth hard spheres. More generally, it is possible to include the effects of rough spheres (Lun 1991) and hard, non-spherical shapes but that will not be considered here.

The forces between hard spheres are singular (impulsive) and so the formal collision operator in eqn. (10) must be changed accordingly to (Brey, Dufty, & Santos 1997; Dufty & Baskaran 2005; Ernst 2000; Lutsko 2004; van Noije & Ernst 2001)

$$C_i(\mathbf{r}_1, \mathbf{v}_1 / \{f_i(t)\}) = -\sum_{j=1}^{s} \int d\mathbf{r}_2 d\mathbf{v}_2 \overline{T}_{ij}(\mathbf{r}_{12}, \mathbf{v}_{12}) \mathscr{F}_{ij}(\mathbf{r}_1, \mathbf{v}_1; \mathbf{r}_2, \mathbf{v}_2 / \{f_k(t)\}) \tag{38}$$

The action of the force has been replaced by a binary scattering operator $\overline{T}_{ij}(\mathbf{r}_{12}, \mathbf{v}_{12})$

$$\overline{T}_{ij}(\mathbf{r}_{12}, \mathbf{v}_{12}) = \sigma_{ij}^2 \int d\hat{\sigma}\,\Theta(\hat{\sigma}\cdot\mathbf{g}_{12})(\hat{\sigma}\cdot\mathbf{g}_{12})\left(\alpha_{ij}^{-2}\delta(\mathbf{r}_2 - \mathbf{r}_1 + \boldsymbol{\sigma})b_{ij}^{-1} - \delta(\mathbf{r}_2 - \mathbf{r}_1 - \boldsymbol{\sigma})\right). \tag{39}$$

The delta functions represent the fact that velocity changes only for particles at contact and on the hemisphere with particles directed at each other. Further details of the dynamics of hard particles are not required here and the interested reader is referred to the literature (see for example and references, Appendix A of Baskaran, Dufty, & Brey 2008).

The "revised" Enskog kinetic theory results from an approximation to $\mathscr{F}_{ij}(\mathbf{r}_1, \mathbf{v}_1; \mathbf{r}_2, \mathbf{v}_2 / \{f_k(t)\})$ as follows,

$$\mathscr{F}_{ij}(\mathbf{r}_1, \mathbf{v}_1; \mathbf{r}_2, \mathbf{v}_2 / \{f_k(t)\}) = \chi_{ij}(\mathbf{r}_1, \mathbf{r}_2 / \{n_k\}) f_i(\mathbf{r}_1, \mathbf{v}_1, t) f_j(\mathbf{r}_2, \mathbf{v}_2, t), \tag{40}$$

where $\{n_k\}$ are the nonequilibrium densities (i.e., the velocity integrals of $\{f_k\}$). To interpret this approximation, consider the integral of $\mathscr{F}_{ij}(\mathbf{r}_1,\mathbf{v}_1;\mathbf{r}_2,\mathbf{v}_2 / \{f_k(t)\})$ over all velocities

$$\int d\mathbf{v}_1 d\mathbf{v}_2 \mathscr{F}_{ij}(\mathbf{r}_1,\mathbf{v}_1;\mathbf{r}_2,\mathbf{v}_2 / \{f_k(t)\}) = \int d\mathbf{v}_1 d\mathbf{v}_2 f_{ij}(\mathbf{r}_1,\mathbf{v}_1;\mathbf{r}_2,\mathbf{v}_2,t)$$
$$= n_i(\mathbf{r}_1,t) n_j(\mathbf{r}_2,t) g_{ij}(\mathbf{r}_1,\mathbf{r}_2,t) \qquad (41)$$

which defines the nonequilibrium pair distribution function for positions, $g_{ij}(\mathbf{r}_1,\mathbf{r}_2,t)$. This is being replaced in the Enskog approximation by a universal functional of the nonequilibrium density $\chi_{ij}(\mathbf{r}_1,\mathbf{r}_2 / \{n_k\})$. For a molecular fluid the functional is fixed by the requirement that $g_{ij}(\mathbf{r}_1,\mathbf{r}_2,t)$ should be the known functional at equilibrium. A similar requirement can be imposed here by the HCS solution (Lutsko 2001, 2002). Finally, it is noted that while the above Enskog approximation retains correlations of positions it completely neglects velocity correlations. This is consistent with the equilibrium state for a molecular fluid, but not for nonequilibrium states in general. For a granular fluid it does not apply exactly even for the HCS, and so it is an uncontrolled assumption here that the effects of such velocity correlations are relatively small for many properties of interest.

The revised Enskog equation now follows from eqns. (9), (38), and (39)

$$\left(\partial_t + \mathbf{v}_1 \cdot \nabla_{\mathbf{r}_1}\right) f_i(\mathbf{r}_1,\mathbf{v}_1;t) + \nabla_{\mathbf{v}_1} \cdot \left(m_i^{-1} F_{0i}(\mathbf{r}_1,\mathbf{v}_1) f_i(\mathbf{r}_1,\mathbf{v}_1;t)\right) = C_i^E(\mathbf{r}_1,\mathbf{v}_1 | \{f_k(t)\}), \qquad (42)$$

with the collision operator

$$C_i^E(\mathbf{r}_1,\mathbf{v}_1 | \{f_k(t)\}) \equiv -\sum_{j=1}^{s} \sigma^2{}_{ij} \int d\mathbf{v}_2 \int d\hat{\sigma}\, \Theta(\hat{\sigma}\cdot\mathbf{g}_{12})(\hat{\sigma}\cdot\mathbf{g}_{12})$$
$$\times [\chi_{ij}(\mathbf{r}_1,\mathbf{r}_1-\boldsymbol{\sigma} | \{n_k\}) f_i(\mathbf{r}_1,\mathbf{v}''_1;t) f_j(\mathbf{r}_1-\boldsymbol{\sigma},\mathbf{v}''_2;t)$$
$$- \chi_{ij}(\mathbf{r}_1,\mathbf{r}_1+\boldsymbol{\sigma} | \{n_k\}) f_i(\mathbf{r}_1,\mathbf{v}_1;t) f_j(\mathbf{r}_1+\boldsymbol{\sigma},\mathbf{v}_2;t)]. \qquad (43)$$

Here $\mathbf{v}''_\alpha \equiv b_{ij}^{-1}\mathbf{v}_\alpha$ represents the velocities due to restituting collisions (the inverse of eqn. (36)). It is easily seen that the granular Boltzmann kinetic equation (Brey, Dufty, Kim, Santos 1998) is recovered in the low density limit for which $\chi_{ij} \to 1$ and $f_j(\mathbf{r}_1 \pm \boldsymbol{\sigma}_{ij},\mathbf{v}_2;t) \to f_j(\mathbf{r}_1,\mathbf{v}_2;t)$ since $\sigma_{ij}$ is small compared to the mean free path. At higher densities both the effects of $\chi_{ij} \neq 1$ and the delocalization of colliding pairs become important, and are included in the revised Enskog theory. As a historical note, the original Enskog equation has $\chi_{ij}$ as a constant (Ferziger & Kaper 1972) rather than a functional of $\{n_i\}$, the actual nonequilibrium densities (van Beijeren & Ernst 1973, 1979). While the earlier version is adequate to describe transport in a one component fluid, it is inconsistent with nonequilibrium thermodynamics for mixture transport (López de Haro, Cohen & Kincaid 1983). The revised theory resolves this problem, and also provides an exact description of the equilibrium fluid and solid phases for hard spheres (Kirkpatrick, Das, Ernst & Piasecki 1990). Thus the revised kinetic equation has the capacity to describe complex states of fluids with solid-like clusters which can occur more commonly in granular matter. Applications to molecular fluids suggests that predictions have quantitative accuracy up to moderately dense states ($n\sigma^3 \leq 0.2$) and semi-quantitative accuracy for some properties up to the freezing density (Alley, Alder, & Yip 1983; Boon & Yip 1991). Similar accuracy is observed for granular fluids as well, but further conditioned by the degree of inelasticity.

In summary, the revised Enskog kinetic theory is a remarkably simple yet accurate description of the complex dynamics under conditions where many-body effects are expected to be important. Theoretical descriptions beyond the Enskog approximation are necessary for very dense, glassy metastable states for molecular fluids (for early references see (Alley, Alder, & Yip 1983; Boon & Yip 1991; Dorfman & Kirkpatrick 1980; Sjogren 1980; Alley, Alder, & Yip 1983)). Its qualitative limitations for granular fluids are still under investigation.

## HCS Solution

For spatially homogeneous states, the normal solution to the Enskog equation is the HCS and has the form

$$f_{hi}(\mathbf{v},\{n_k\},T_h(t)) = n_i v_0^{-3} \phi_i\left(\frac{\mathbf{v}-\mathbf{U}_h}{v_0(T_h(t))};\{n_h\}\right). \tag{44}$$

where $v_0(T_h) = \sqrt{2T_h/m}$ is a "thermal velocity" associated with the temperature $T_h$ and $m = \sum_{i=1}^{s} m_i/s$ is the average mass. This defines the dimensionless distribution $\phi_i$. The uniform densities $\{n_{hi}\}$, uniform temperature $T_h(t)$, and flow velocity $\mathbf{U}_h$ in eqn. (44) are those for the HCS. This scaling property is special to the hard sphere interaction and follows from dimensional analysis since there is no intrinsic energy scale in this case. The functional forms of the HCS distributions are obtained from eqn. (35)

$$-\frac{1}{2}\zeta_h \nabla_\mathbf{v} \cdot (\mathbf{V} f_{hi}) = \sum_j \chi_{ij} J_{ij}^{(0)}(V \mid f_{hi}), \tag{45}$$

and $J_{ij}^{(0)}(\mathbf{v}_1 \mid f_{hi})$ are the Boltzmann collision operators for a low density granular mixture

$$J_{ij}^{(0)}(\mathbf{v}_1 \mid f_{hi}) \equiv \sigma_{ij}^2 \int d\mathbf{v}_2 \int d\hat{\sigma}\,\Theta(\hat{\sigma}\cdot\mathbf{g}_{12})(\hat{\sigma}\cdot\mathbf{g}_{12})[\alpha_{ij}^{-2} f_i(\mathbf{v}_1'',t) f_j(\mathbf{v}_2'',t) - f_i(\mathbf{v}_1,t) f_j(\mathbf{v}_2,t)]. \tag{46}$$

(the low density Boltmann operators occurs here because for the local HCS the Enskog and Boltzmann operators differ only by the factors $\chi_{ij}$). Use has been made of translational and rotational invariance to reduce $\chi_{ij}(\mathbf{r}_1,\mathbf{r}_1 - \boldsymbol{\sigma}_{ij} \mid \{n_k\})$ to a constant function of $\{n_i\}$, denoted simply by $\chi_{ij}$. Aside from normalization, all other density dependence of the $\{f_{hi}\}$ occurs through the $\{\chi_{ij}\}$. In the elastic limit, $\alpha_{ij} \to 1$, the solutions are all Maxwellians at the constant temperature $T$, as required. Otherwise the $f_{hi}$ have a quite different functional form.

It was noted above that the partial temperatures for each species do not constitute hydrodynamic fields, but rather are functions of $T$ and $\{n_i\}$. Still, they may be useful properties to characterize the solutions to eqn. (45). The partial temperature $T_i$ is related to the species kinetic energy analogous to eqn. (5), and the associated cooling rate is defined by $T_i^{-1}\partial_t T_i = -\zeta_i$. Multiplying eqn. (45) by $m_i V^2$ and integrating gives $\zeta = \zeta_i$ for all species $i$. Thus, in the HCS the common feature among species is their cooling rates rather than their temperatures. In fact, it can be shown that for mechanically different species the partial temperatures are different (Garzó & Dufty 1999). Equipartition of energies is a property of equilibrium states, implying a common temperature for all degrees of freedom, but violated for nonequilibrium states including all granular states. This has been verified directly via MD simulation (Dahl, Hrenya, Garzo, Dufty 2002).

## Methods of Solution

The focus of this chapter is the derivation hydrodynamic equations from the kinetic equation. As described above, this requires a special "normal solution" to the kinetic equation in which all space and time dependence occurs through the hydrodynamic fields. The explicit construction of such a solution for the Enskog kinetic equation is the subject of the next section. However, it is appropriate to digress for a brief overview of some other, more general, methods to solve the Enskog equation. Such solutions provide a more detailed representation of the fluid from macroscopic length and time scales down to those for the particles. Although more numerically intensive in applications than hydrodynamics, these solutions are appropriate when validity conditions for a continuum description fail.

The most accurate method for solution to the Enskog equation is that of direct simulation Monte Carlo (DSMC). Originally developed to solve the Boltzmann equation (Bird 1994, Garcia & Alexander 1997), it has been extended to the Enskog equation as well (Montanero & Santos 1996) and has a direct analog for the granular fluid. This is a simple and efficient algorithm applicable to a wide range of conditions.

In principle, the solution to the kinetic equation is equivalent to specifying all moments of the distribution function. Equations for the moments are obtained by taking moments of the Enskog equation, leading to a hierarchy of equations coupling lower order moments to those of higher order. Approximate solutions are obtained by truncating this hierarchy. This method was first applied to the Boltzmann equation in Grad 1960 and is known as the Grad moment method. An early application of moment methods to two-phase flows was given in Simonin 1996. A generalized, more accurate form of the Grad moment method for the Enskog equation has been given in Lutsko 1997 and Lutsko 2004 and applied to granular shear flow far outside the Navier-Stokes domain. A related method based on information from low order moments is the Maximum Entropy method (Koopman 1969), currently being applied to polydisperse gas-solid flows.

The next section is devoted to finding the normal solution to the Enskog equation according to the Chapman-Enskog method. This method does not address the most general normal solution, but rather its form when the spatial gradients of the hydrodynamic fields are sufficiently small to allow a perturbation expansion around a local HCS. The extension of the Chapman-Enskog method to arbitrary reference states has been proposed by Lutsko 2006, but will not be considered here.

## Chapman - Enskog Solution

Now consider the general case of a spatially inhomogeneous state and return to the problem of finding the normal solution. The functional dependence of the normal distribution function on the hydrodynamic fields at all points can be given an equivalent local representation in terms of the fields and all their derivatives at a given point

$$f_i(\mathbf{r}, \mathbf{v} | \{y_\alpha(t)\}) = f_i(\mathbf{v}, \{y_\beta(\mathbf{r},t), \nabla_\mathbf{r} y_\beta(\mathbf{r},t),...\}). \tag{47}$$

If the system is only weakly inhomogeneous a further Taylor series expansion can be carried out to give

$$f_i(\mathbf{v}, \{y_\beta(\mathbf{r},t), \nabla_\mathbf{r} y_\beta(\mathbf{r},t),...\}) = f_i^{(0)}(\mathbf{v}, \{y_\beta(\mathbf{r},t)\}) + f_i^{(1)}(\mathbf{v}, \{y_\beta(\mathbf{r},t), \nabla_\mathbf{r} y_\beta(\mathbf{r},t)\}) + .... \tag{48}$$

Here, $f_i^{(0)}$ depends on the fields but not the gradients, and $f_i^{(1)}$ is both a function of the fields and linear in their gradients. The dots in the above equation denote second and higher powers of $\nabla_\mathbf{r} y_\alpha$ as well as higher degree derivatives of $y_\alpha$. This is an expansion in the small spatial variations of the hydrodynamic fields over distances of the order of the mean free path $\ell_{mfp} |\nabla_\mathbf{r} y_\alpha| / y_\alpha$. Here, the mean free path is defined as $\ell_{mfp} = 1/n\sigma^2$, where $n$ is a characteristic number density and $\sigma$ is a characteristic particle size. The Chapman-Enskog method is a procedure for constructing a normal solution to a given kinetic equation as an expansion in these small spatial gradients (Ferziger & Kaper 1972). Formally, a solution of the form eqn. (48) is substituted into the kinetic equation and terms of each order in the small parameter are set equal to zero to determine the solution. There are two complications in ordering the kinetic equation. The first is the choice for any "size" of external force as measured in terms of the small parameter, and the second is the size of the time derivatives for each term in the distribution function. The scaling of the force depends on the conditions of interest, such as a gravitational or other external field, such as the coupling to the gas phase in gas-solid flows, whose effect on transport may be weak or strong when compared to that due to spatial nonuniformities in the fields. A proper assessment of this force with respect to all dimensionless parameters of both the gas and solid phases is essential for a correct modeling of gas-solid flows and the justification of the two fluid model resulting from the Chapman-Enskog solution. In Garzo, Dufty, & Hrenya 2007 the magnitude of the force was arbitrarily taken to be of first order in the expansion parameter (e.g., of first order in the Knudsen number defined by the length scale for variation of the hydrodynamic fields). Here, however, to illustrate the Chapman-Enskog method in its simplest form the analysis is restricted to the case for which no the external forces are present.

The ordering of the time derivative is accomplished by noting that the time derivatives of $f_i(\mathbf{r}, \mathbf{v} | \{y_\alpha(t)\})$ are proportional to the time derivatives of the fields, since the solution is normal. Furthermore, it is required that these hydrodynamic fields be solutions to the balance equations. The latter relate the time derivatives of the fields to their spatial gradients, providing the desired ordering of the time derivatives,

$$\partial_t y_\alpha(t) = \partial_t^{(0)} y_\alpha(t) + \partial_t^{(1)} y_\alpha(t) + \ldots, \tag{49}$$

For example, substituting directly the expansion (49) into eqns. (21) – (23) and identifying contributions to each order in the gradients gives for the first and second order derivatives

$$\partial_t^{(0)} n_i = 0, \quad \partial_t^{(0)} T = -\zeta_h^{(0)} T, \quad \partial_t^{(0)} U_\beta = 0, \tag{50}$$

and

$$\partial_t^{(1)} n_i + \nabla \cdot (n_i \mathbf{U}) + m_i^{-1} \nabla \cdot \mathbf{j}_{0i}^{(0)} = 0, \tag{51}$$

$$\partial_t^{(1)} T = -\frac{3}{2n}\left(P_{\gamma\beta}^{(0)} \partial_{r_\gamma} U_\beta - \nabla \cdot \mathbf{q}^{(0)}\right) + \sum_{i=1}^{s}\frac{T}{nm_i} \nabla \cdot \mathbf{j}_{0i}^{(0)} - \zeta_h^{(1)} T, \tag{52}$$

$$\rho \partial_t^{(1)} U_\beta = -\rho \nabla \cdot \mathbf{U} - \partial_{r_\gamma} P_{\gamma\beta}^{(0)}. \tag{53}$$

The fluxes and cooling rate with superscripts $0$ or $1$ denote the result obtained from eqns. (27) – (30) using $f_i^{(0)}$ or $f_i^{(1)}$, respectively. Note that these are not equations for the fields, which are not being expanded. Rather they are definitions of $\partial_t^{(0)} y_\alpha(t)$ and $\partial_t^{(1)} y_\alpha(t)$ in terms of the fields and their gradients. The fact that $\partial_t^{(0)} \neq 0$ (due to collisional cooling) is a new feature of granular fluids. Its exact incorporation in the Chapman-Enskog expansion assures that the small gradient expansion places no restrictions on the degree of inelasticity.

It is now straightforward to implement the Chapman-Enskog procedure to determine $f_i^{(0)}$ and $f_i^{(1)}$ in eqn. (48). The details are described elsewhere (for recent reviews and lists of early references see (Garzo, Dufty, Hrenya 2007; Lutsko 2005) and only the results described here. At lowest order the Enskog kinetic equation is of the same form as the HCS equation eqn. (45). Hence, its solution, $f_i^{(0)}$ is the *local* HCS distribution

$$f_i^{(0)} = n_i v_0^{-3}(T) \phi_i \left( \frac{|\mathbf{v} - \mathbf{U}|}{v_0(T)}; \{n_k\} \right). \tag{54}$$

More specifically, it is the same functional form as in eqn. (44) except with $\{n_{hi}\}$, $T_h(t)$, and $\mathbf{U}_h$ replaced by the actual nonequilibrium fields $\{n_i(\mathbf{r},t)\}$, $T(\mathbf{r},t)$, and $\mathbf{U}(\mathbf{r},t)$. It should be emphasized that this leading order "reference state" is not chosen as the basis for the expansion, but rather is a consequence of the Chapman - Enskog procedure. There is no flexibility to choose a different reference state for the chosen small gradient expansion.

To first order in the gradients the solution is

$$f_i^{(1)} = \sum_m \mathcal{H}_i^m(\mathbf{V}; T, \{n_k\}) \cdot \nabla_\mathbf{r} y_m. \tag{55}$$

Here $\{y_m\}$ is a set of independent linear combinations of $\{n_i\}, T$ and components of $\mathbf{U}$. They can be chosen such that the linear equations determining the coefficients $\mathcal{H}_i^m(\mathbf{V})$ all have the form

$$\left( (\mathcal{L} - \lambda^{(m)}) \mathcal{H}^m \right)_i = A_i^m. \tag{56}$$

The linear operator $\mathcal{L}$ is defined in terms of the linearized forms of the Enskog operators in eqn. (46)

$$(\mathcal{L} X)_i = \frac{1}{2} \zeta^{(0)} \nabla_\mathbf{v} \cdot (\mathbf{V} X)_i + (LX)_i, \tag{57}$$

$$(LX)_i = -\sum_{j=1}^s \chi_{ij} \left( J_{ij}^{(0)}[\mathbf{v}_1 / X_i, f_j^{(0)}] + J_{ij}^{(0)}[\mathbf{v}_1 / f_j^{(0)}, X_i] \right). \tag{58}$$

The right sides of eqn. (56), $\mathcal{H}_i^m$, are explicit functions determined from $\{f_i^{(0)}(t)\}$ and $\mathcal{F}_{ij}(\mathbf{r}_1, \mathbf{v}_1; \mathbf{r}_2, \mathbf{v}_2 / \{f_k^{(0)}(t)\})$ for the Enskog form eqn. (40). Hence they are known from the lowest order solution. Finally, $\lambda^{(m)} \Leftrightarrow \left( 0, \frac{1}{2}\zeta^{(0)}, -\frac{1}{2}\zeta^{(0)}, -\frac{1}{2}\zeta^{(0)}, -\frac{1}{2}\zeta^{(0)} \right)$ are the smallest eigenvalues of $\mathcal{L}$.

From the corresponding eigenvectors and their properties, it is possible to show that the integral equations (56) have solutions and that they are unique (Garzó & Dufty 1999; Dufty, Garzo, Hrenya 2007).
In summary, the exact normal solution to the generalized Enskog equation has been constructed up through terms to first order in the gradients of the hydrodynamic fields. In particular, the first two terms eqn. (54) and eqn. (55) together with the equations that determine them, eqn. (45) and eqn. (56) place no limitations on the degree of inelasticity or density, beyond those implied by the kinetic equation itself.

The results of this subsection require $\ell_{mfp} |\nabla_{\mathbf{r}} y_\alpha| / y_\alpha \ll 1$. For finite geometries this implicitly requires $\ell_{mfp} \ll D$, where $D$ is any relevant system dimension. For domains where this Knudsen number is not small this small gradient solution fails and the constitutive equations of the next section do not apply. In some cases this can be corrected by modified boundary conditions (slip conditions). Otherwise, the solution does not apply and a different solution to the kinetic equation must be constructed.

## Constitutive Equations

The constitutive equations required for hydrodynamics are obtained from eqns. (27) – (30), specialized to the case of smooth, inelastic hard spheres. Substitution of the Chapman-Enskog solution eqn. (54) and eqn. (55) into these expressions gives the constitutive equations to first order in the gradients. There are two types of contributions, arising from the first terms on the right sides evaluated with $f_i^{(0)}$ and $f_i^{(1)}$, and those obtained from evaluating the two particle functional $\mathcal{F}_{ij}(\mathbf{r}_1, \mathbf{v}_1; \mathbf{r}_2, \mathbf{v}_2 / \{f_k(t)\})$ to first order in the potential. The latter are called "collisional transfer" contributions. They vanish at low densities, but dominate at high densities. Included in the collision transfer contributions are first order gradient terms from products $f_i^{(0)} f_j^{(0)}$ for a colliding pair depending on the fields due to the different positions of the colliding pair at $\mathbf{r}$ and $\mathbf{r} \pm \boldsymbol{\sigma}_{ij}$. Also, there are collisional transfer contributions from the functional dependence of $\chi_{ij}(\mathbf{r}_1, \mathbf{r}_1 \pm \boldsymbol{\sigma}_{ij} / \{n_k\})$ on the species densities.

There are many simplifications due to the assumption of fluid symmetry. For example, to zeroth order in the gradients the mass and energy fluxes vanish, while the cooling rate and pressure tensor become

$$\zeta^{(0)} = \frac{\pi}{12nT} \sum_{i,j} \left(1 - \alpha_{ij}^2\right) \frac{m_i m_j}{m_i + m_j} \chi_{ij}^{(0)} \sigma_{ij} \int d\mathbf{v}_1 \int d\mathbf{v}_2 f_i^{(0)}(V_1) f_j^{(0)}(V_2) / |\mathbf{V}_1 - \mathbf{V}_2|^3, \tag{59}$$

$$P_{\gamma\beta}^{(0)} = p\delta_{\gamma\beta} = \left[ nT + \frac{2}{9}\pi \sum_{i,j} \left(1 + \alpha_{ij}\right) \frac{m_i m_j}{m_i + m_j} \chi_{ij}^{(0)} \sigma_{ij}^3 n_j \int d\mathbf{W}^2 f_i^{(0)}(V) \right] \delta_{\gamma\beta}. \tag{60}$$

Recall that the $\{f_i^{(0)}\}$ are local HCS solutions so that the fields in these equations are the physical values $\{T(\mathbf{r},t), \{n_i(\mathbf{r},t)\}\}$ to be determined from the resulting hydrodynamic equations.

The complete set of constitutive equations to first order in the gradients is

$$\zeta \to \zeta^{(0)} + \zeta_U \nabla \cdot \mathbf{U}, \tag{61}$$

$$\mathbf{j}_{0i} \to -\sum_{j=1}^{s} m_i m_j \frac{n_j}{\rho} D_{ij} \nabla \ln n_j - \rho D_i^T \nabla \ln T, \tag{62}$$

$$\mathbf{q} \to -\lambda \nabla T - \sum_{i,j=1}^{s} T^2 D_{q,ij} \nabla \ln n_j, \tag{63}$$

$$P_{\beta\mu} \to p\delta_{\beta\mu} - \eta\left(\partial_\beta U_\mu + \partial_\mu U_\beta - \frac{2}{3}\delta_{\beta\mu}\nabla \cdot \mathbf{U}\right) - \kappa\delta_{\beta\mu}\nabla \cdot \mathbf{U}. \tag{64}$$

These expressions are characterized by transport coefficients that are functions of the temperature and the species densities, determined from integrals over the solutions to the integral equations eqn. (56). There are $(s-1)(2s+1)$ transport coefficients for the mass flux, $D_{ij}$ and $D_i^T$, the shear and bulk viscosity for the pressure tensor, $\eta$ and $\kappa$, and $2s^2+1$ coefficients for the heat flux, $D_{q,ij}$ and $\lambda$. There remain only the practical issues of solving the equations (45) and (56), which are discussed below.

## Approximate Evaluation

The above constitutive equations, and definitions of the transport coefficients, are direct consequences of the generalized Enskog equation. Any limitations must be attributed to approximations leading to that kinetic equation since the Chapman-Enskog method gives an exact construction (to each order in the perturbation expansion considered). Applications of these results require the explicit solutions to eqn. (45) for the HCS distribution, $f_i^{(0)}$, and to equations (56) for the functions $A^{(m)}$ to determine the transport coefficients. In the following approximation methods for each are described briefly.

Consider first the solution to eqn. (45) by representing it as an expansion in a complete set of associated Laguerre polynomials (Sonine polynomials, up to a normalization constant) (Ferziger & Kaper 1972). First write eqn. (54) as

$$f_i^{(0)} = n_i v_0^{-3}(T)\phi_i(c;\{n_k\}); \quad \mathbf{c} = \frac{\mathbf{v}-\mathbf{U}}{v_0(T)}, \tag{65}$$

$$\phi_i(c;\{n_k\}) = \phi_{Mi}(c;\{n_k\})\left[1+\sum_{n=2}^{\infty} b_n L_n^{1/2}(c^2)\right]. \tag{66}$$

Here $\phi_{Mi}(c;\{n_k\})$ is a Maxwellian chosen to have the exact second moment of $\phi_i(c;\{n_k\})$. Hence it is parameterized by the kinetic temperature $T_i$ which must be determined self-consistently from the condition $\zeta = \zeta_i$ (see discussion of the HCS above). An approximate solution is obtained by truncating the expansion. The coefficients are determined by substitution of the truncated expansion into eqn. (45) and taking moments of that equation with respect to all polynomials retained in the truncation. There is a complication in this case due to the nonlinearity of eqn. (45) which implies that the truncated series is not sufficient to determine the truncated set of coefficients $\{b_n\}$. Consequently, there is no unique way to determine their best estimate from a given truncation. This situation has been reviewed recently in Santos & J. Montanero 2008. Fortunately, the simplest truncation at $n = 2$ and a linearization of eqn. (45) for $b_2$ gives a very good approximation (compared to Monte Carlo solution to the kinetic equations) except at very strong dissipation (Brey, Ruiz-Montero & Cubero 1996; van Noije & Ernst 1998).

A similar method applies for the solution to the linear integral equations (56). The solutions are given a series representation as above

$$\mathscr{A}_i^m(\mathbf{c}) = \phi_{Mi}(c;\{n_i\})\left[\sum_{n=1}^{\infty} a_n^m L_n^{1/2+\lambda_m}(c^2)\right]. \tag{67}$$

The parameter $\lambda_m$ takes the value 0, 1 or 2 depending on whether $A_i^m$ is a scalar, vector, or second order tensor. Again, truncation of the series defines a particular approximation. Substitution into eqn. (56) and taking the corresponding moments now determines the coefficients of that truncation exactly, since the equations are linear. Again, leading order truncations give very good approximations except at strong dissipation (Montanero, Santos & Garzó 2005). Recently, it has been shown that higher order truncations extend the accuracy to strong dissipation (Garzo, Santos & Montanero 2007, Garzo, Vega Reyes & Montanero 2008).

An instructive test of this approximation scheme and its results is given by a recent study of impurity diffusion (Garzo & Vega Reyes 2009). The above Enskog analysis is specialized to a binary mixture in which one of the species has very small concentration, denoted with a subscript $0$. The constitutive equation (62) becomes

$$\mathbf{j}_{00} = -\frac{m_0^2}{\rho} D_0 \nabla n_0 - \frac{m_0 m}{\rho} D \nabla n - \rho D^T \nabla \ln T. \tag{68}$$

Consider, for example, the diffusion coefficient $D_0$. The corresponding expression from eqn. (27) is

$$D_0 = -\frac{\rho}{3 m_0 n_0} \int d\mathbf{v} \mathbf{V} \cdot \mathscr{A}_0(\mathbf{V}), \tag{69}$$

where $\mathscr{A}_0$ is the solution to an integral equation of the type eqn. (56)

$$\mathscr{L}\mathscr{A}_0 = -\mathbf{V} f_0^{(0)}. \tag{70}$$

The first two terms in the expansion of $\mathbf{A}_0$ are

$$\mathscr{A}_0(\mathbf{V}) = f_M(V)[a_1 \mathbf{V} + a_2 \mathbf{S}(V) + ...], \tag{71}$$

where $\mathbf{S}(V) = \mathbf{V}\left(\frac{1}{2} m_0 V^2 - \frac{5}{2} T_0\right)$. Then substitution of eqn. (71) into eqn. (70) and taking moments with respect to $\mathbf{V}$ and $\mathbf{S}(V)$ gives the two coupled equations for $a_1$ and $a_2$. Finally, with these known $D_0$ is determined from eqn. (69). When there is no strong mechanical dissimilarity of the impurity from the host fluid, the first Sonine approximation gives excellent agreement with DSMC simulations of the Enskog equation. However, the second Sonine approximation is required when there is significant mechanical difference (see the last section below).

To summarize this section on Navier-Stokes hydrodynamics for the hard sphere fluid, the central approximation has been the choice of the generalized Enskog kinetic theory (as opposed to the simpler Boltzmann equation or more complex high density phenomenological equations). As for normal fluids, quantitative predictions of this kinetic theory appear to be limited to low or moderately dense gases. This depends on the property being calculated, with good results for pressure up to solid densities while those for transport coefficients begin to fail for volume fractions $\phi = \pi n \sigma^3 / 6 > 0.1$. The domain of accuracy for granular fluids is further compromised as the degree of inelasticity increases, $\alpha < 0.8$. Nevertheless, the Enskog kinetic theory remains semi-quantitative over a much wider range of densities and inelasticities. The derivation of the normal solution to this kinetic equation to obtain the Navier-Stokes equations does not involve further approximation. Rather it requires verification of experimental conditions for small hydrodynamic variations; the form and coefficients of the first order Chapman-Enskog expansion are given exactly within the Enskog kinetic theory for all densities and degrees of inelasticity. The practical evaluation of these coefficients may involve a further approximation, such as truncation of a series representation. Therefore, comparison of experimental results to Navier-Stokes hydrodynamics first entails assurance that the conditions for small gradients are met. Having done so any observed discrepancies can be attributed to a failure to implement accurately the consequences of the kinetic equation (evaluation of the integral equations) or to the kinetic equation itself.

## COMPLEX FLUID HYDRODYNAMICS

As emphasized in the comment following eqns. (27) - (30) the concept of a macroscopic, hydrodynamic description is more general than the Navier-Stokes equations. This is well-known for high molecular weight fluids that exhibit a wealth of rheological effects not present at the Navier-Stokes level. The surprising and interesting feature of granular fluids is that such effects occur commonly even for structurally simple systems such as hard spheres. For such states, their hydrodynamic description requires a more complete determination of the normal solution eqn. (24), beyond the leading order (Navier-Stokes) Chapman-Enskog expansion of the last section. The resulting hydrodynamics will be referred to as "complex fluid" hydrodynamics.

The simplest example is the Burnett hydrodynamics obtained by retaining the first two terms of the Chapman-Enskog expansion (the results for weak inelasticity are given by Sela & Goldhirsch 1998, and Kumaran 2004). Even at this level new physical effects occur, such as non-zero viscometric functions and complex heat flow. However, in most cases for which the Navier-Stokes conditions fail it is not sufficient to add small corrections (e.g., higher order Chapman-Enskog approximations). Instead, more complex constitutive equations that are highly nonlinear in the gradients can be expected. Examples are granular fluids under shear, with constitutive equations that are complex functions of the shear rate.

For simple molecular systems complex fluid behavior is rare since the system must be driven hard externally to generate the necessary large gradients. Granular fluids are different because, in addition to external driving forces (e.g., the coupling to a gas phase), there is the internal mechanism of collisional energy loss. The latter is particularly effective in the formation of steady states that would not be possible for molecular fluids. Consider a fluid between two parallel plates, each at the same fixed temperature. A normal fluid has its steady state at a uniform temperature, while that for a granular fluid has a temperature that is lower in the middle than at the edges due to collisional cooling. Thus the granular fluid "generates" gradients on its own, independent of applied external forces. These inherent gradients cannot necessarily be made small in order to assure Navier-Stokes hydrodynamics (Hrenya, Galvin & Wildman 2008; Galvin, Hrenya & Wildman 2007).

A second example is the same configuration but with the wall in relative parallel motion to shear the fluid. The shear does work creating viscous heating. For the molecular fluid this is compensated either by heating the uniform fluid, or by generating a steady state with non-uniform temperature gradient. However, the granular fluid can support a steady, uniform state with viscous heating compensated by collisional cooling. In this latter case the system autonomously seeks a steady state temperature that is determined by the imposed shear rate, $a$, and the degree of inelasticity, $\alpha$. The dimensionless gradient in this case is $a^* = a/\nu$ where $\nu$ is a characteristic collision frequency that scales for hard spheres as the square root of the steady temperature. Any attempt to make $a^*$ small enough to justify the Navier-Stokes conditions by decreasing $a$ has the effect of decreasing the steady state temperature in just such a way that $a^*$ remains unchanged. The system again sets its own gradient, and in this case it has been shown (Santos, Garzo & Dufty 2004) that the Navier-Stokes description of uniform shear is never justified and the fluid is inherently complex. Nevertheless, a complex fluid hydrodynamics may apply. The Enskog kinetic theory has been applied to the derivation of constitutive equations in special cases, such as the state of uniform shear flow and polydisperse conditions, with significant success (Lutsko 2004; Montanero, Garzo, Alam & Luding 2006). Unfortunately, universal features of constitutive equations for classes of complex fluid states have not been identified and may not exist.

In addition to these complexities due to the internal mechanism of collisional energy loss for granular fluids, the Navier-Stokes equations may fail in gas-solid flows due to the local driving forces from the gas phase. When these drive the formation of bubbles or dense plugs there can be large density and thermal gradients for which the Navier-Stokes constitutive equations are not justified. In this case a more general normal solution to the Enskog equation itself must be considered. This remains an open area of current research.

# DISCUSSION

In this chapter, the conceptual basis of granular hydrodynamics as arising from an underlying kinetic theory has been reviewed. First, in the most general context, the definition of a kinetic theory was given. Then, using the exact balance equations for the fields of interest and a normal solution to the kinetic equation, the constitutive equations were obtained expressing the fluxes and the source in terms of the fields, and hence a closed hydrodynamic description for the fields was obtained. Finally, the analysis was specialized to the theoretically tractable setting of smooth inelastic hard spheres and the kinetic theory approximated by the revised Enskog equation. In this setting, the constitutive equations were obtained explicitly and exactly to first order in the spatial gradients of the fields for the case of a weakly inhomogeneous system. The resulting hydrodynamic equations constitute the granular Navier-Stokes hydrodynamics, with the pressure, cooling rate, and transport coefficients defined as averages over solutions to linear integral equations. Practical application of these results based on the first Sonine approximation appear in reference (Garzó & Dufty 1999) for a one component fluid, and in (Garzo, Dufty, & Hrenya 2007) for multicomponent systems. These results apply for arbitrary dissipation and are extensions of the earlier work of Jenkins (Jenkins 1998), and of Jenkins and Mancini (Jenkins & Mancini 1989), respectively, which made the additional approximation of weak dissipation. The reader is referred to the literature for the explicit expressions for the parameters of the Navier-Stokes equations.

As noted at the outset, one of the primary advantages of deriving the hydrodynamic equations systematically is that the approximations made at every step are clearly stated and the validity of the resulting theory can be tested in unambiguous terms. In the remainder of this closing section examples of some numerical tests are presented.

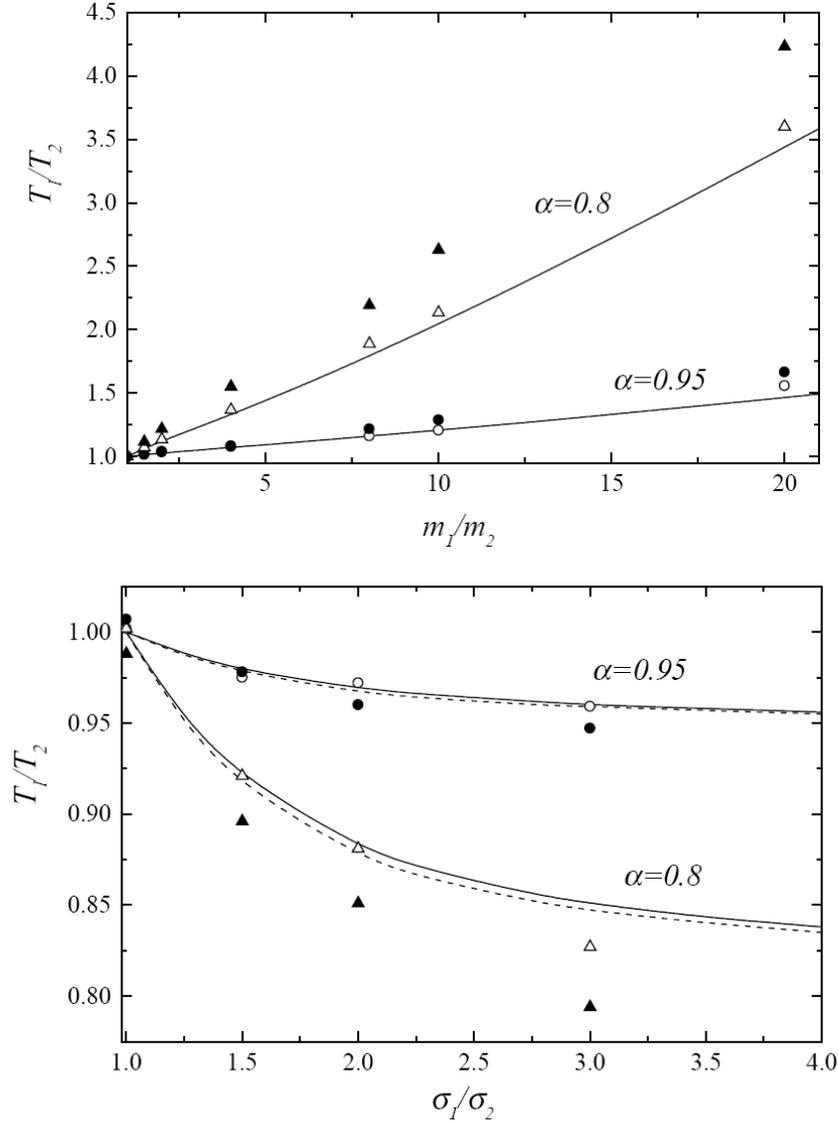

*Figure 1. Plot of the temperature ratio $T_1/T_2$ as a function of a) Top Panel: the mass ratio $m_1/m_2$ for $\sigma_1/\sigma_2 = 1$ and b) Bottom Panel: the size ratio for $m_1/m_2 = 1$. $\phi_1 = \phi_2 \equiv \phi$ in both panels. The symbols are MD simulations and the lines are Enskog predictions. Two different values of $\alpha$, $\alpha = 0.95$ (circles) and $\alpha = 0.8$ (triangles) and two different values of $\phi$, $\phi = 0.1$ (open symbols, solid line) and $\phi = 0.2$ (solid symbols, dashed line) are shown (reproduced from Dahl, Hrenya, Garzo, Dufty 2002) © [2002][American Physical Society]. Used with permission.*

Once the specialization to smooth, inelastic hard spheres was made, the subsequent derivation here included three approximations at various stages in its development. The first is the assumption that the Enskog kinetic equation (42), is a reliable mesoscopic description of a moderately dense system of inelastic hard spheres. This can be tested by molecular dynamics (MD) simulation of Newton's equations of motion to compute properties of interest for comparison of their calculation from the Enskog equation. The singular inelastic hard sphere dynamics is implemented using an event-driven algorithm in which the particles traverse straight lines until a pair is at contact; the velocities of the pair are then changed according to the collision rules eqn. (36). These same collision rules are used in the Enskog equation which can be solved numerically with no further approximations needed, using an extension of the Direct Simulation Monte Carlo (DSMC) method developed for the Boltzmann equation and extended to the Enksog equation (Montanero & Santos 1996). A comparison of the results of these two numerical studies provides a means to quantify the domain of validity for the Enskog equation. Of course, this will depend somewhat on the property and state considered. Figure 1 shows the comparison from (Dahl, Hrenya, Garzo, Dufty 2002) of the temperature ratio for a binary mixture in its HCS as a function of mass ratio (left panel) and size ratio (right panel), from MD simulation (symbols) and a calculation based on the Enskog equation (lines). Two values for the restitution coefficient are shown, $\alpha = 0.95$ and $\alpha = 0.8$, and two values of volume fractions, $\phi = 0.1$ and $\phi = 0.2$, where $\phi = \pi n \sigma^3 / 6$. It is seen that the Enskog theory is accurate for both values of $\alpha$ for $\phi = 0.1$ over a wide range of mass and size ratios. The results are less accurate for $\phi = 0.2$, but still quite good for the $\alpha = 0.95$ case. Figure 2 shows a comparison of the temperature ratio and the shear viscosity $\eta$ as a function of $\alpha$ for a granular fluid undergoing uniform shear flow with a mass ratio of 4 and $\phi = 0.1$ ( Montanero, Garzo, Alam, Luding 2006). There is excellent agreement for the shear viscosity for the range $\alpha \geq 0.7$ considered; this degree of accuracy extends to the overall temperature and pressure as well (not shown). However, the accuracy of Enskog for the temperature ratio decreases as the degree of dissipation increases.

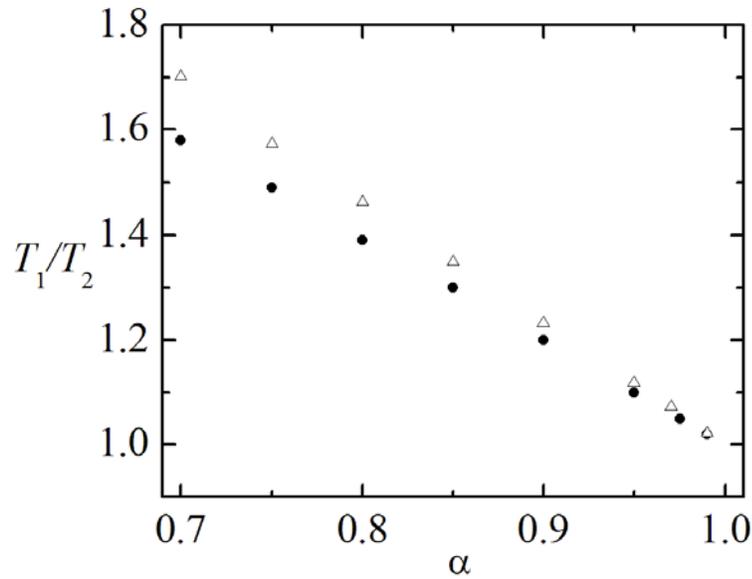

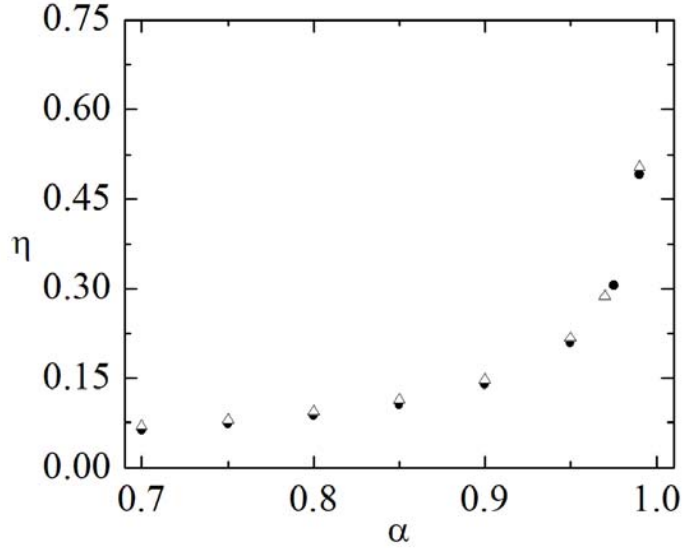

*Figure 2. Shear viscosity $\eta$, and temperature ratio $T_1/T_2$ versus the coefficient of $\alpha$ with $\phi_1/\phi_2 = \sigma_1/\sigma_2 = 1$, $m_1/m_2 = 4$, and $\phi = 0.1$. The solid circles correspond to the DSMC results and open triangles correspond to MD simulations (adapted from Montanero, Garzo, Alam, Luding 2006 ).*

      Consider next the second approximation made in the derivation presented here, namely the construction of explicit solutions to eqns. (45) and (56) using a truncated Sonine polynomial expansion. A chosen truncation can be tested by comparing properties calculated using the analytic results to those obtained from DSMC simulations. First, consider the Sonine expansion as applied to the HCS distribution $f_h$ for a one component fluid. In this case the HCS solutions to the dimensionless Enskog and Boltzmann equations are the same. DSMC studies were first performed in reference (Brey, Ruiz-Montero & Cubero 1996) and extended in (Brilliantov & Pöschel 2006). It is found that the leading Sonine approximation is quite accurate for $\alpha > 0.6$ except for very large velocities. For $\alpha < 0.6$ and for large velocities at any $\alpha < 1$ this approximation fails. Next, consider the approximate solution to the integral equations that determine the various transport coefficients. Figure 3 shows the shear viscosity of a one component granular fluid as a function of $\alpha$ and as a function of $\phi$ from DSMC (symbols) and from the first Sonine approximation (lines) (Montanero, Santos, & Garzó 2005). The agreement is quite good over the entire parameter space of density and dissipation. Conditions under which this leading approximation fails is illustrated in Figure 4 for the case of impurity diffusion described above (Garzo &Vega Reyes 2009). The impurity diffusion coefficient $D_0$ is shown as a function of $\alpha$ at $\phi = 0.2$ as determined from DSMC (symbols), from the first Sonine approximation (dashed line), and from the second Sonine approximation (solid line). The top panel is for an impurity that is heavier and larger than the host fluid particles, while the bottom panel is for a lighter and smaller impurity. There is excellent agreement between all three calculations in the former case, while the first Sonine approximation fails badly in the second case. Thus, mechanical differences among constituent particles can require higher order approximations.

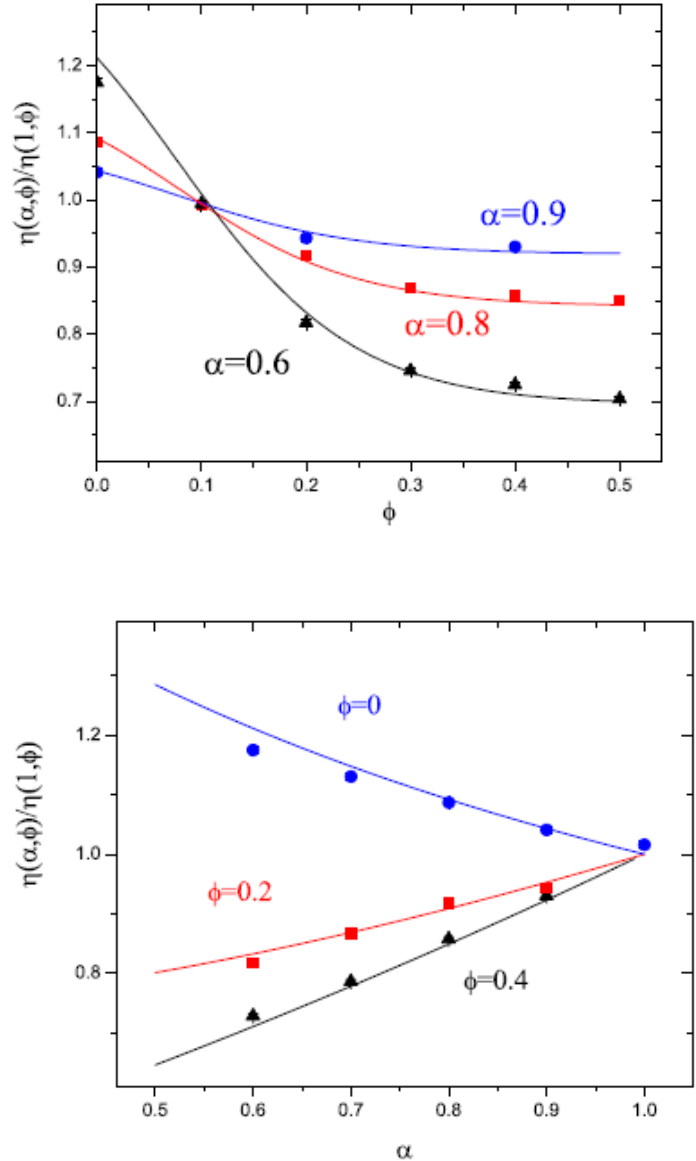

*Figure 3. Shear viscosity $\eta$ of a one component granular fluid a) as a function of $\alpha$ for different values of $\phi$ and b) as a function of $\phi$ for different values of $\alpha$. The symbols correspond to the DSMC results and the lines correspond to the first Sonine approximation (reproduced from Montanero, Santos and Garzo 2005) © [2005][American Institute of Physics]. Used with permission.*

    The third approximation is the restriction to inhomogeneous states with weak spatial gradients of the hydrodynamic fields, in order to justify truncation of the Chapman-Enskog expansion. For molecular fluids this is a matter of controlling the experimental conditions responsible for the inhomogeneity (sufficiently weak initial perturbations, weak boundary forces). It has been demonstrated that this can be done for granular systems as well in many cases, justifying a Navier-Stokes descriptions. For example, it has been found that experimental hydrodynamic profiles of a supersonic granular flow past a wedge are

described well by the Enskog Navier-Stokes hydrodynamic equations (Rericha, Bizon, Shattuck, Swinney 2002). Similarly, the hydrodynamic profiles of a three dimensional vibrated granular system are well described by these same hydrodynamic equations, when sufficiently far away from the vibrating wall (Huan et al 2004). These and other experimental studies suggest that there is a wide variety of experimental conditions that can be reliably captured by a hydrodynamic description using the Navier-Stokes constitutive equations.

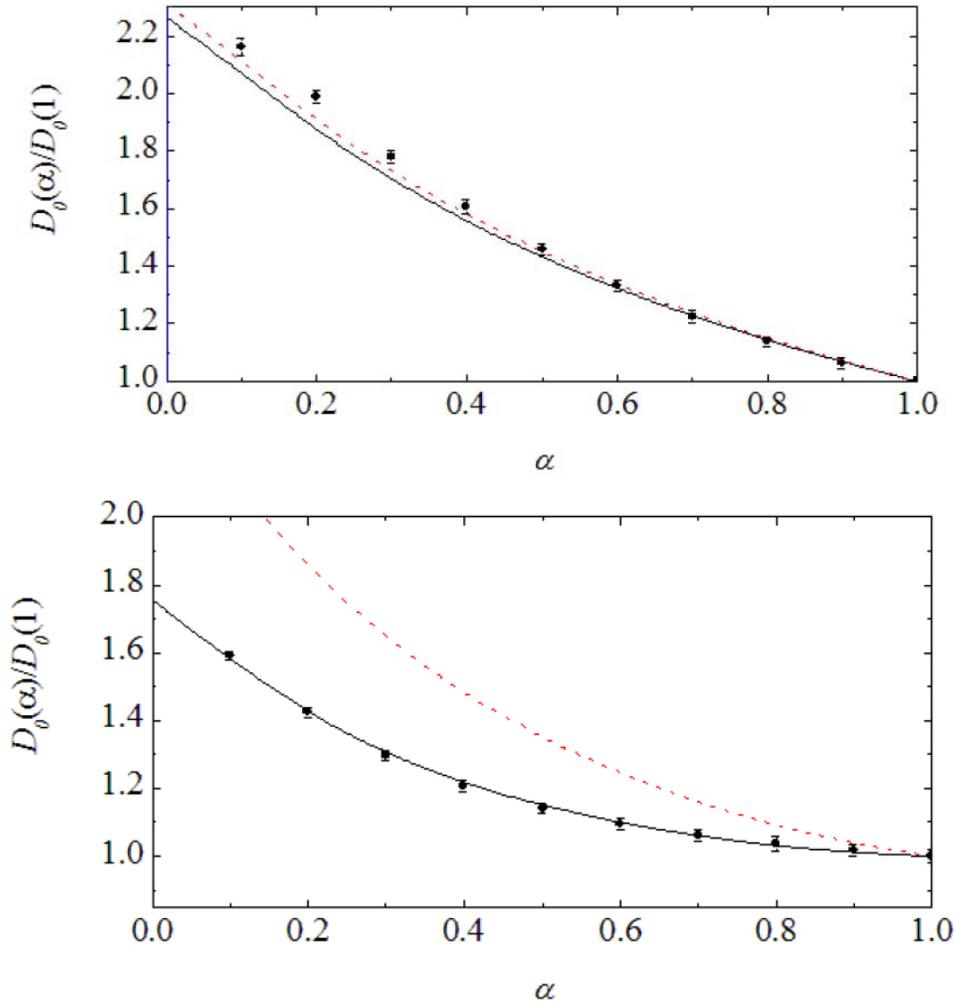

*Figure 4. Plot of the reduced diffusion coefficient $D_0(\alpha)/D_0(1)$ as a function of the coefficient of restitution for $\phi = 0.2$, with $m_0/m = 2$, $\sigma_0/\sigma = 2$ in the top graph and $m_0/m = 1/5$, $\sigma_0/\sigma = 2$ in the bottom graph. The solid lines correspond to the second Sonine approximation and the dashed lines refer to the first Sonine approximation. The symbols are the results obtained from DSMC (Garzo & Vega Reyes 2009) © [2009][American Physical Society]. Used with permission*

     As noted in the previous section, granular fluids also support states for which the gradients cannot be controlled externally and some higher order, more complex hydrodynamics is required. Recent examples have been discussed in reference (Galvin, Hrenya, Wildman 2007; Hrenya, Galvin, Wildman 2008) for both simulation and experiment. This raises the question of whether the Enskog kinetic equation also provides an adequate description for more complex states with large gradients and extreme conditions. Figure 5 illustrates results for a polydisperse (continuous distribution of size, mass, and

restitution coefficient) granular fluid under shear (Lutsko 2004). The reduced shear rate $a^*$, which scales as the inverse root steady state temperature, is shown as a function of an average restitution coefficient $\alpha_0$ for three densities $n^* = n\sigma^3$. As described in the previous section, for a given shear rate and restitution coefficient the system seeks a steady state by balancing viscous heating and collisional cooling. This figure shows that the DSMC simulation of the Enskog equation gives excellent agreement with MD simulations. Also shown are the results of an approximate solution to the Enskog equation based on moments of the distribution. This is not hydrodynamics in the sense described here, but closely related to it. The rheological properties of the pressure tensor are also well-described by the Enskog equation for this complex granular fluid under conditions far from the Navier-Stokes weak gradient restrictions.

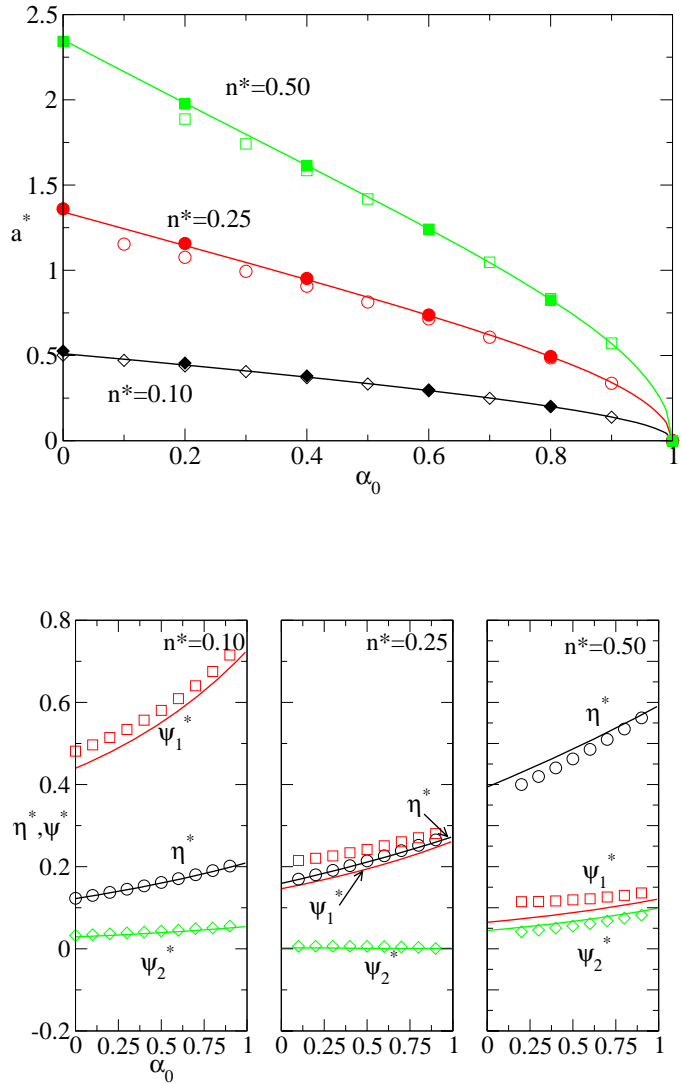

Figure 5. The top figure shows the reduced shear rate $a^*$ as a function of $\alpha_0$ as determined from the DSMC (filled symbols), and MD (open symbols) for three values of the reduced density $n^* = n\sigma^3$. Also shown is a generalized moment solution to the Enskog equation, GME (lines). The lower three figures show the reduced shear viscosity $\eta^*$ and the two viscometric functions $\psi_1$ and $\psi_2$ for each of the

*densities. Symbols are from MD and lines are from GME (reproduced from Lutsko 2004)* © *[2004][American Physical Society]. Used with permission.*

In summary, there are qualitative conditions of dissipation $0.8 \leq \alpha \leq 1$ and packing fractions $0 < \phi < 0.2$ for which the Navier-Stokes hydrodynamic description, with coefficients evaluated in a low order Sonine approximation, can be expected to be reliable, if the experimental conditions verify the restriction to weak gradients. When this Navier-Stokes description does fail, it could be attributed to the breakdown of any of the three approximations above, i.e., 1) The system is in a state such that the assumption of weak gradients does not hold and hence the constitutive equations are more complex (flow down an inclined plane belongs in this category); 2) The coefficient of restitution could lie in a region where the polynomial approximation breaks down higher order terms in the expansions are required; or 3) the Enskog kinetic theory itself fails due to velocity correlations that have a quantitative importance for the properties of interest. Although the above examples give some indication, a more complete identification of the domain of validity for the Enskog description and its Navier-Stokes predictions with approximate evaluations of the associated coefficients is an area of ongoing investigation in the granular physics community via theory, simulation, and experiment. Such studies also present a fertile set of questions that need to be answered to move granular hydrodynamics to its full potential beyond both Navier-Stokes and/or the Enskog kinetic theory.

The derivation of Navier-Stokes hydrodynamics described here requires an approximation for the kinetic theory at an early stage. There is an alternative route to this same hydrodynamics, resulting in formally exact expressions for the constants in these equations. This is an approach based in statistical mechanics and leads, for example, to Green-Kubo time correlation function expressions for the transport coefficients. One method to evaluate these coefficients is by approximate kinetic theories. The advantage of this approach is that such approximations are postponed to the last stage. These well-developed linear response methods for molecular fluids have been extended recently to granular fluids (Baskaran, Dufty & Brey 2008; Dufty 2009; Dufty, Baskaran, & Brey 2008), and it has been shown that an evaluation of the Green-Kubo expressions via an approximate kinetic theory neglecting velocity correlations leads to the same results as those obtained here (Baskaran, Dufty & Brey 2007). These two approaches provide complementary starting points for exploration of effects beyond the Enskog approximation that are required at higher densities.

## AKNOWLEDGEMENT

The authors are grateful to V. Garzo, C. Hrenya, J. Lutsko, and A. Santos for providing their figures and associated data. AB acknowledges support by the NSF through Grants DMR- 0705105 and DMR- 0806511.

## NOMENCLATURE

| | |
|---|---|
| $f_i(\mathbf{r},\mathbf{v};t)$ | number density of species $i$ at $\mathbf{r},\mathbf{v},t$. |
| $n_i(r,t)$ | number density of species $i$ at $\mathbf{r},t$. |
| $e(\mathbf{r},t)$ | energy density |
| $\mathbf{p}(\mathbf{r},t)$ | momentum density |
| $T(\mathbf{r},t)$ | granular temperature |
| $\mathbf{U}(\mathbf{r},t)$ | flow velocity |
| $C_i(\mathbf{r}_1,\mathbf{v}_1 \mid \{f_k(t)\})$ | collision operator for species density $i$ |
| $w(\mathbf{r},t)$ | energy density loss rate |
| $\mathbf{j}_i(\mathbf{r},t)$ | mass flux for species $i$ |
| $\mathbf{j}_{0i}(\mathbf{r},t)$ | mass diffusion flux for species $i$ |

| Symbol | Description |
|---|---|
| $\mathbf{s}(\mathbf{r},t)$ | total energy flux |
| $\mathbf{q}(\mathbf{r},t)$ | total heat flux |
| $t_{\gamma\beta}(\mathbf{r},t)$ | total momentum flux tensor |
| $P_{\beta\gamma}(\mathbf{r},t)$ | pressure tensor |
| $\mu_{ij} = m_i/(m_i + m_j)$ | reduced mass for the pair $i,j$ |
| $\zeta(\{n_i\},T)$ | cooling rate |
| $\alpha_{ij}$ | restitution coefficient for hard sphere collision of species $i,j$ |
| $m_i, \sigma_i$ | particle mass and diameter for species $i$ |
| $\mathcal{F}_{ij}(\mathbf{r}_1,\mathbf{v}_1;\mathbf{r}_2,\mathbf{v}_2\mid\{f_k\})$ | two particle functional for kinetic equation "closure" |
| $C_i^E(\mathbf{r}_1,\mathbf{v}_1\mid\{f_k\})$ | Enskog collision operator |